\documentclass{emulateapj}

\shortauthors{Straughn et al.}

\usepackage{subfigure}
\usepackage{longtable}

\newcommand{\etal}	{\mbox{et al.\,}}

\newcommand{\Mo}	{\mbox{M$_{\odot}$}}

\newcommand{\cge}	{\ensuremath{\gtrsim}}

\newcommand{\Hline}[1]{\mbox{H{\footnotesize {#1}}}}
\newcommand{\Ha}{\Hline{\mbox{$\alpha$}}\thinspace}
\newcommand{\Hb}{\Hline{\mbox{$\beta$}}\thinspace}

\newcommand{\OII}{[O\,{\sc ii}]}
\newcommand{\OIII}{[O\,{\sc iii}\thinspace]}
\newcommand{\CIV}{C\,{\sc iv}\thinspace}
\newcommand{\CIII}{C\,{\sc iii}\thinspace]}
\newcommand{\fluxsv}{$\times$ $10^{-17}$ ergs cm$^{-2}$ s$^{-1}$}
\newcommand{\fluxei}{$\times$ $10^{-18}$ ergs cm$^{-2}$ s$^{-1}$}

\begin{document}

\title{Emission--Line Galaxies from the HST Probing Evolution And Reionization Spectroscopically (PEARS) Grism Survey I: The South Fields}

\author{Amber N. Straughn\altaffilmark{1,2}, Norbert Pirzkal\altaffilmark{3}, Gerhardt R. Meurer\altaffilmark{4}, Seth H. Cohen\altaffilmark{5}, Rogier A. Windhorst\altaffilmark{5}, Sangeeta Malhotra\altaffilmark{5}, James Rhoads\altaffilmark{5}, Jonathan P. Gardner\altaffilmark{2}, Nimish P. Hathi\altaffilmark{6}, Rolf A. Jansen\altaffilmark{5}, Norman Grogin\altaffilmark{3}, Nino Panagia\altaffilmark{3}, Sperello di Serego Alighieri\altaffilmark{7}, Caryl Gronwall\altaffilmark{8}, Jeremy Walsh\altaffilmark{9}, Anna Pasquali\altaffilmark{10}, Chun Xu\altaffilmark{11}}

\altaffiltext{1}{Amber.N.Straughn@nasa.gov}
\altaffiltext{2}{Astrophysics Science Division, Observational Cosmology Laboratory, Goddard Space Flight Center, Code 665, Greenbelt, MD 20771}
\altaffiltext{3}{Space Telescope Science Institute, Baltimore, MD 21218}
\altaffiltext{4}{Department of Physics and Astronomy, Johns Hopkins University, Baltimore, MD 21218}
\altaffiltext{5}{School of Earth and Space Exploration, Arizona State University, Tempe, AZ 85287}
\altaffiltext{6}{Department of Physics \& Astronomy, University of California, Riverside, CA 92521}
\altaffiltext{7}{INAF - Osservatorio Astrofisico di Arcetri, I-50125 Firenze,
Italy}
\altaffiltext{8}{Department of Astronomy \& Astrophysics, Pennsylvania State University, University Park, PA, 16802, USA}
\altaffiltext{9}{ESO Space Telescope European Co-ordinating Facility, D-85748 Garching bei München, Germany}
\altaffiltext{10}{Max--Planck--Institut for Astronomie, Koenigstuhl 17, D-69117 Heidelberg, Germany}
\altaffiltext{11}{Shanghai Institute of Technical Physics, 200083 Shanghai, China}


\begin{abstract}
We present results of a search for emission--line galaxies in the
Southern Fields of the Hubble Space Telescope PEARS (Probing Evolution
And Reionization Spectroscopically) grism survey.  The PEARS South
Fields consist of five ACS pointings (including the Hubble Ultra Deep
Field) with the G800L grism for a total of 120 orbits, revealing
thousands of faint object spectra in the GOODS-South region of the
sky.  Emission--line galaxies (ELGs) are one subset of objects that
are prevalent among the grism spectra.  Using a 2-dimensional
detection and extraction procedure, we find 320 emission lines
orginating from 226 galaxy ``knots'' within 192 individual galaxies.
Line identification results in 118 new grism--spectroscopic redshifts
for galaxies in the GOODS-South Field.  We measure emission line
fluxes using standard Gaussian fitting techniques.  At the resolution
of the grism data, the \Hb\ and \OIII\ doublet are blended.  However,
by fitting two Gaussian components to the \Hb\ and \OIII\ features, we
find that many of the PEARS ELGs have high \OIII/\Hb\ ratios compared
to other galaxy samples of comparable luminosities.  The
star--formation rates (SFRs) of the ELGs are presented, as well as
a sample of distinct giant star--forming regions at z$\sim$0.1--0.5
across individual galaxies.  We find that the radial distances of
these HII regions in general reside near the galaxies' optical
continuum half--light radii, similar to those of giant HII regions in
local galaxies.

\end{abstract}

\section{Introduction}
The Probing Evolution And Reionization Spectroscopically
(PEARS~\footnote{http://archive.stsci.edu/prepds/pears}) ACS grism survey
provides a 200 HST orbit dataset from which one can investigate many
different aspects of galaxy evolution.   High-redshift objects
such as Ly$\alpha$ galaxies, Lyman break galaxies, and AGN are being investigated by Rhoads \etal (2009) and Grogin \etal (2009, in preparation).  Elliptical galaxies (Ferreras \etal 2009, in preparation), and emission--line galaxies (ELGs;
Straughn \etal 2008) are also being studied.   A similar deep grism program
was carried out in the GRAPES project (Pirzkal \etal 2004, Xu \etal
2007).  Here we discuss results of a search for
ELGs in the PEARS South Fields.  In particular, we present new grism
spectroscopic redshifts for 118 galaxies in the GOODS South Field, as
well as discuss the ELG line luminosities, star--formation rates, and AGN
candidates among the sample.

For many years, galaxies that are actively forming stars have been regarded as important objects to study in the context of galaxy assembly.  In particular, the \Ha, \OIII, and \OII\ lines have been used extensively to determine their SFRs (Kennicutt 1983; Gallego \etal 1995; Gallego \etal 2002; Brinchmann \etal 2004; Westra \& Jones 2007; Kewley \etal 2004; Glazebrook \etal 2004).  Many projects have specifically used slitless spectroscopy in order to study ELGs.  Ground-based slitless spectroscopy has been used by Kurk \etal (2004) to identify ELGs.  Yan \etal (1999) derived the H$\alpha$ luminosity function and SFR for galaxies at z$\cge$1 using the HST NICMOS G141L grism from the NICMOS Grism Parallel Survey (McCarthy \etal 1999).  Teplitz \etal (2003) studied ELGs using the STIS Parallel Survey (Gardner \etal 1998) and Drozdovsky \etal (2005) presented ELGs from the HST ACS Grism Parallel Survey.  Shim \etal (2009) have studied the luminosity function and evolution of the SFR density for ELGs using H$\alpha$, also using the NICMOS data.  The GRism ACS Program for Extragalactic Science (GRAPES; Pirzkal \etal 2004, Malhotra \etal 2005) has also yielded slitless spectroscopy for galaxies in the Hubble Ultra Deep Field (HUDF), including a large sample of ELGs (Pirzkal \etal 2006, Xu \etal 2007).  PEARS is a follow-up grism survey to GRAPES, and provides a larger spectroscopic dataset of ELGs in an 8$\times$ larger area.  In Straughn \etal (2008) we investigated in detail several methods aimed at detecting these ELGs in the PEARS HUDF pointing.  In the current paper we use the most efficient method and extend that study to include the remaining four PEARS South ACS Fields.  In Section 2 we discuss the PEARS dataset used here.  Section 3 outlines the methods used to detect the ELGs.  In Section 4 we present results of the search, including a table of the South Field ELGs detected along with new spectroscopic redshifts, and a discussion of line luminosities, star-formation rates, AGN candidates, and the radial distribution of galaxy knots.  In Section 5 we summarize our findings and discuss future prospects.

\section{Data}
The HST PEARS grism survey consists of nine ACS Fields observed with
the G800L grism.  The G800L grism yields low-resolution
($R\!\sim\!100$) optical spectroscopy between $\lambda =$6000-9500{\AA}.  Five
fields were observed in the GOODS South region (including the Hubble
Ultra Deep Field) and four in GOODS North.  Here we present properties
of ELGs detected in the PEARS South fields.  The PEARS HUDF was
observed for 40 orbits (four roll angles, obtaining spectra for
sources with limiting continuum AB magnitude
$i'_{AB}\!\lesssim\!27.0$~mag).  The other four South PEARS fields were observed for
20 orbits each (three roll angles per field with spectra for sources with
limiting continuum AB magnitude $i'_{AB}\!\lesssim\!26.0$~mag).  Limiting continuum magnitudes are estimated from the net spectral significance, which is a measure of the peak integrated S/N of a stacked spectrum, and is described fully for the GRAPES project in Pirzkal \etal (2004; see also Cohen, S. \etal 2009, in preparation).
Observations at multiple roll angles were made in each field in order to reduce the
contamination from overlapping spectra in crowded regions.  These
multiple roll angles are also used in detecting viable
emission--line sources, as described in the following section.
Malhotra \etal (2009, in preparation) will describe the PEARS ACS grism observations
in detail.  Pirzkal \etal (2004) give a detailed description of the
closely--related prior GRAPES project.  The PEARS North Fields are currently
being reduced and a future paper presenting ELGs from the North Fields will be Paper II in this series of PEARS ELGs studies.

\section{Methods}
We briefly outline the procedures used to detect ELGs in the PEARS grism
data, using a 2D--detection method that takes advantage of the
observation that emission lines typically originate from clumpy knots
of star formation within galaxies.  A detailed description of this method and
comparison with several other extraction methods are given in Straughn \etal
(2008).

\subsection{Data Pre-Processing}
The first step in the grism data reduction involves pre--processing of
the grism data.  Each grism image is median filtered and smoothed
using a 13 x 3 smoothing kernel along the direction of the dispersion
axis (i.e. unsharp-masked).  We refer to Meurer \etal (2007) for a full
description of this method of pre--processing ACS grism data in
general.  The dimension of smoothing kernal used does not greatly affect the sources that are selected.  The choice of 13x3 smoothing kernel ensures efficient detection of real emission-line objects while largely avoiding faint image defects or other contaminants to the sample.  This unsharp--masking step is performed in order to largely remove the
continuum flux from the dispersed image, leaving behind sharp emission
line features.  Zero--order images of compact sources are excluded in
the triangulation step, described in the next section.  Residual image
defects are also retained, but are unique to each roll angle and are
thus excluded in the next steps as described below.  In doing this, we
isolate the actual emission line which would ordinarily be washed out
by the continuum, and therefore missed in more traditional 1D
detection methods (see Figure 1).  After the images
are pre--processed in this manner, they are catalogued with the source
extraction algorithm SExtractor (Bertin \& Arnouts 1996), giving a
list of compact sources.  An average of 820 compact sources are initially
selected from each field in this manner.

\begin{figure}
\centering
\subfigure{
\includegraphics[scale=0.5]{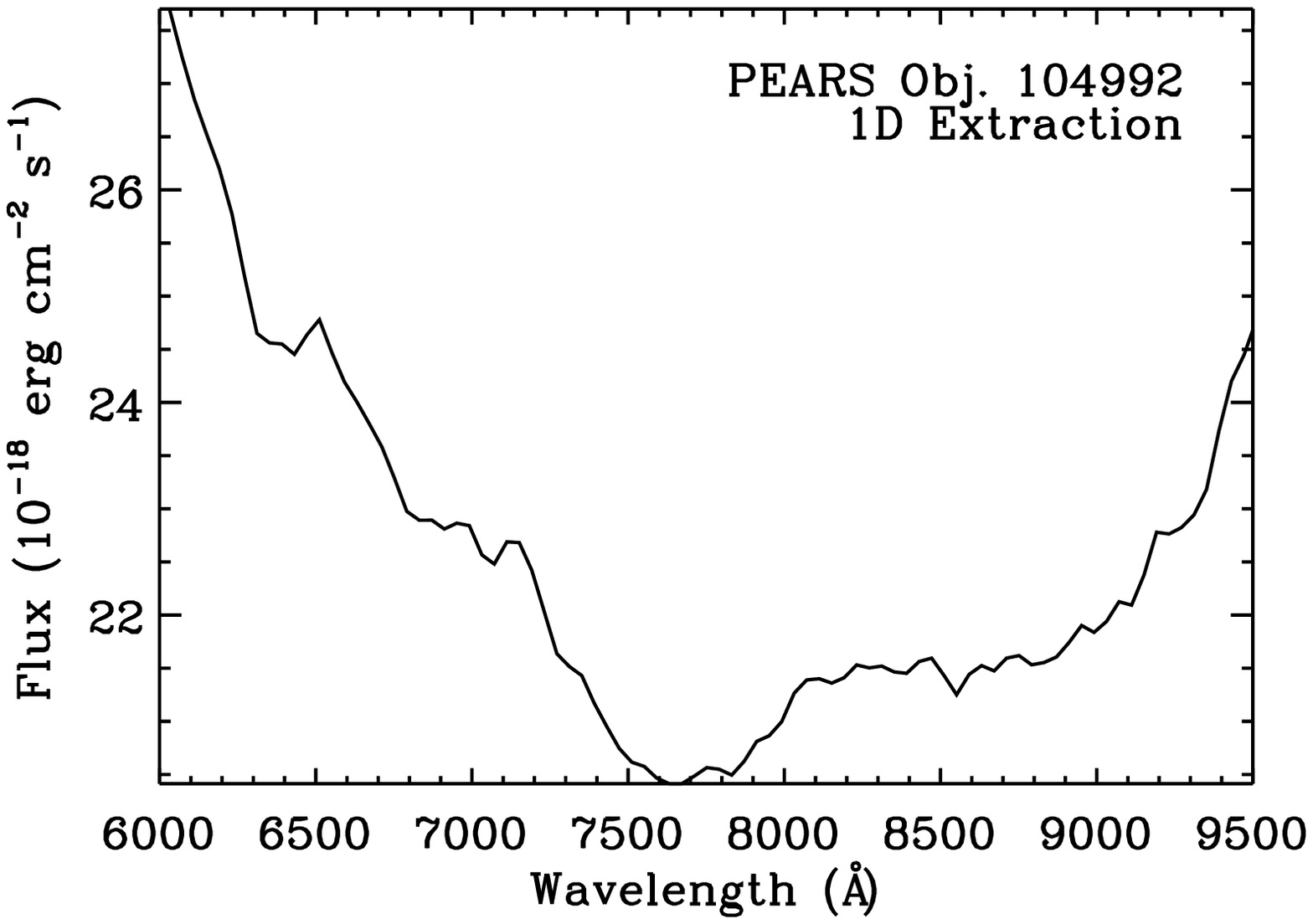}}
\hspace{0cm}
\subfigure{
\includegraphics[scale=0.5]{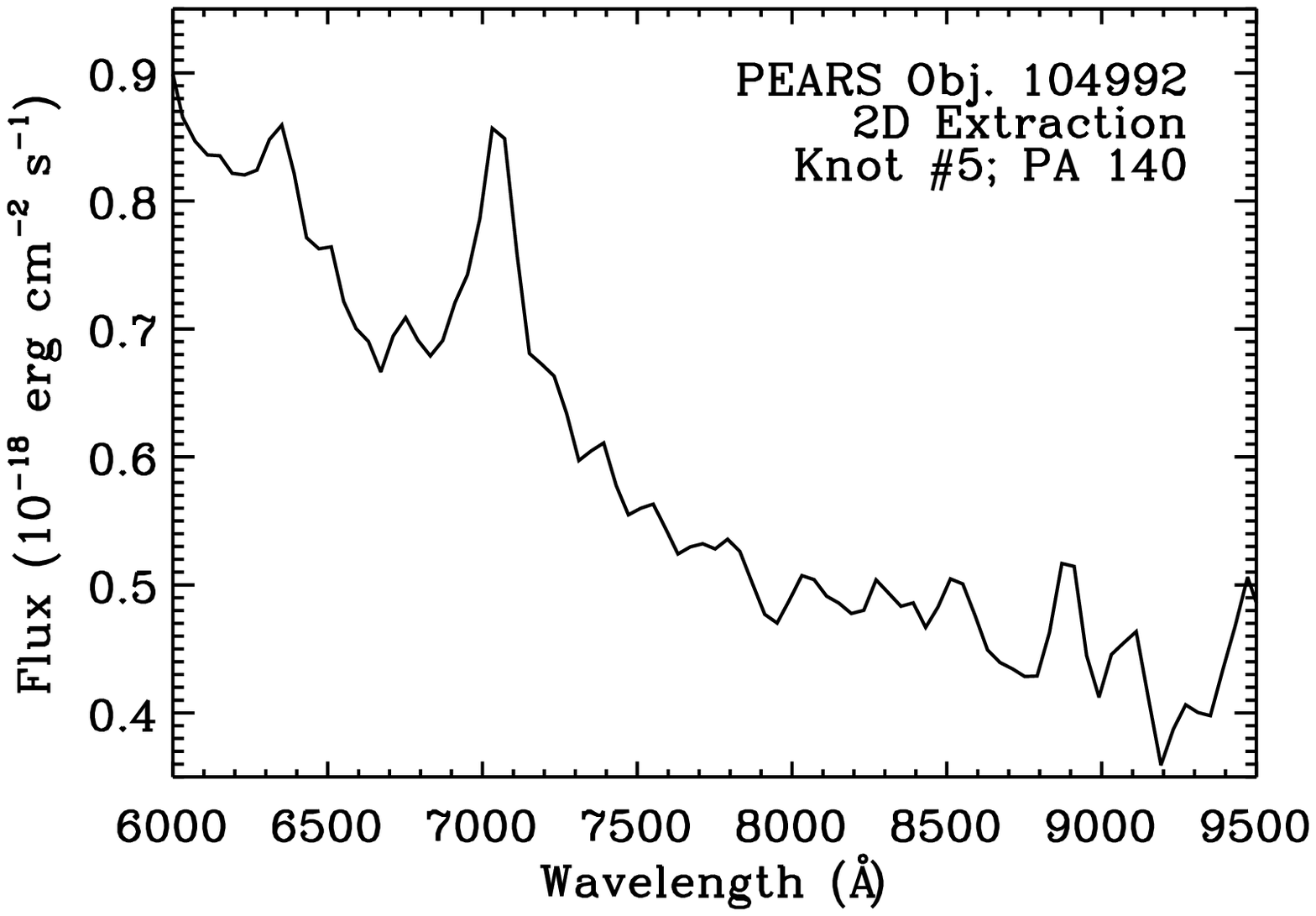}}
\caption{Here we demonstrate the advantage of the 2D--detection method outlined in this paper (and described in detail in Straughn \etal 2008) for PEARS Object 104992.  The top panel shows that continuum flux overwhelms the line when the spectrum of the entire galaxy is extracted (as would be the case in 1D methods; see e.g., Xu \etal 2007).  However, the emission line at (observed-frame) 7000\AA\ is clearly seen when extraction of an individual knot is performed (bottom panel).  See Figure~\ref{fig:multknotsmontage} for an image of this object.}
\label{fig:compspec}
\end{figure}

\subsection{Emission Line Detection by Triangulation}
The basis of this method of 2D emission--line detection and wavelength
calibration relies on each source being observed in more than one roll
angle.  The emitting source is traced back along the dispersion
direction for each roll angle, and intersections of these traces
are used to obtain the real sky coordinates (RA, Dec), as well as the
wavelength solution for that emitting source (see also Figure 2 of Straughn \etal 2008).  In this way, image
defects are excluded from the selection, since they would not
ordinarily appear at the same physical location on the grism images
and map onto a ``source'' as described here.  This procedure is
applied to all roll angle pairs, such that each source---that has
three position angles observed, for example---has three calculations
made (i.e. PA1-PA2, PA1-PA3, PA2-PA3).  The HUDF, which has four
position angles observed, thus has six calculations per source.  This
procedure produces the master catalog of ELG sources, which are then
visually checked.  In this visual confirmation step, there are
occasional instances where an emission--line candidate was present in
all three roll angles, and thus was included in the master catalog,
but is not a genuine emission line.  Such is the case for some bright
galaxies that have continuum ``bumps'' that appear in the grism image
as compact sources: i.e., false line candidates.  When examining the
collapsed 1D spectra from the individual sources, it is clear which
sources are genuine emission lines and which are not.  The genuine
lines are subsequently retained for each field and the final
wavelength for each line listed in Table 1 is obtained by averaging
the results from the roll angle pairs described here.  Here we define
our terminology, since extractions were performed on individual galaxy
``knots'': a galaxy can have several knots, and each knot can have
more than one line as allowed by the grism bandpass.  An average of 90
knot candidates per field are retained in the automated triangulation
step, and an average of 46 genuine knots per field are retained after
the visual confirmation step.  This method produced a total of 320
emission lines originating from 226 galaxy knots, within 192 individual
galaxies in the five total PEARS South Fields.

\subsection{Redshifts of Emission--Line Galaxies}
For ELG knots that have only one emission line in their
spectra---which is the case for 68\% of the galaxy knots---a
first--guess redshift is essential for line identification.  For this we
use the spectroscopic and photometric redshifts from the GOODS--MUSIC
catalog (Grazian \etal 2006 and references therein).  About 33\% of
the ELGs detected in the PEARS South Fields have spectroscopic
redshifts and 85\% have photometric redshifts.  There is almost
complete overlap between the two catalogs---less than 3\% of sources
have spectroscopic redshifts but no photometric redshifts.  Where no
spectroscopic or photometric redshift exists for a particular source,
we match our sources against the table of spectrophotometric redshifts
of Cohen \etal (2009, in preparation)---which are determined by using a combination of 
both the grism spectra and broad--band data (see also Cohen \etal
2009; Ryan \etal 2007).  Spectra with strong lines, however, are often
assigned artificially high spectrophotometric redshifts due to the
presence of such lines that are absent from the template SEDs used.
In total, there were 16 galaxies that had only a spectrophotometric
redshift, five of which had two lines in the observed wavelength
interval and therefore had a grism redshift calculated based on the
line ratio.  Of the galaxies with a single line in the spectrum and
only a spectrophotometric redshift, 3 had spectrophotometric redshifts
in concordance with the observed line and were used to deduce final
identification.  For the 31 objects with a single line where either no
prior redshifts were available, or the spectrophotometric redshifts do
not agree with any of the likely line identifications, no redshift was
assigned.

Line identification proceeds as follows.  For galaxy knots that have both \Ha/\OIII, \OIII/\OII, or \CIII/\CIV in the observed wavelenth range, the ratio of the observed line wavelengths is computed to
obtain a direct line identification and redshift---without need of a
 first--guess redshift.  For galaxy knots with only a single line, the
 existing spectroscopic, photometric, or spectrophotometric redshifts
 (in order of preference) from Grazian \etal (2006) and Cohen \etal
 (2009, in preparation) are used to determine the most likely
 identification of the single line within the redshift and instrinsic
 grism errors.  Redshifts based on these identifications and measured line positions are
 subsequently recalculated and given in Table 1.

 Line fluxes are derived using standard Gaussian fitting techniques
 and measured lines with S/N$\cge$2 are retained in the final catalog
 (Figure 6).  96\% of objects have S/N$\cge$3.  Since
 the \OIII\ line---which is usually the strongest of the lines we
 detect---is blended with \Hb due to the grism spectral resolution, we
 fit two Gaussian components.  In these two--component fits, the
 central wavelengths of the \OIII\ and \Hb\ lines are constrained to
 have the correct wavelength ratio.  In order to reduce the number of
 free parameters that go into the fits of the low--resolution grism spectra, we examine individually a
 subsample of fifteen representative test case spectra, varying the ratio of
 \Hb--to--\OIII\ line widths from 0.1 to 1 (noting that, from the 1D spectra, all \OIII\ line widths are qualitatively
 larger than the weaker, blended \Hb\ line widths).  In these tests,
 we found that an
 average \Hb\ line width of $\sim$0.5 that of the \OIII\ line width
 gave the best quantitative statistical fits.  For 67\% of the
 spectra in which we detect an \OIII\ line, the $\chi^{2}$ improves
 when including the \Hb\ line in our fit.  Of these, 23\% of \Hb\
 lines had $S/N>3$ and were thus included in the final catalog.  Here
 we adopt the higher S/N cutoff ($\cge$3) than for the general catalog
 due to the fact that the line is blended and thus inherently
 contaminated by the \OIII\ doublet, and so only the most secure \Hb\
 lines are included.  In all cases where it was possible to include
 \Hb\ in the line fits---and where such inclusion resulted in improved
 fits---the \Hb\ line was weaker by a factor of at least 2.  Utilizing
 this composite \OIII\ $+$ \Hb\ fitting technique results in 90 \OIII\
 fluxes which are statistically improved using the reduced $\chi^{2}$
 metric, compared to fitting the \OIII\ line alone.  Thirty \Hb\
 fluxes also result from this method.

\section{Results}
In Table 1 we list the emission--line wavelengths, line IDs, fluxes,
and grism redshifts for 320 lines originating from 226 star--forming
knots within 192 individual galaxies found in our search for ELGs in
the PEARS South Fields.  Of these, 25 galaxies (12\%) exhibit multiple emitting
knots, and 61 knots (27.0\%) have two lines (thus providing secure
redshifts; see Section 3).  Our sample includes 136 \OIII, 83 \Ha, 30
\OII, 30 \Hb, 4 \CIV, 3 \CIII, 2 MgII, 1 H$\gamma$, and 1 NeIII lines
(see Table 3).  Of these galaxies, 17 are CDF-S X-ray sources
(Giacconi \etal 2002; Grogin \etal 2009, in preparation).  The most
common lines (\Ha, \OIII, and \OII) are detectable in the redshift
ranges of 0--0.4, 0.1--1.1, and 0.4--1.5 respectively, given the grism
band--pass.  The \OIII\ emitters have, in general, very high
equivalent widths, with a mean rest frame equivalent width
$EW_{[OIII],mean}=152\AA$ at a redshift of z$\sim$0.5.  The equivalent
width distributions of the \Ha, \OIII, and \OII\ lines are shown in
Figure~\ref{fig:ewdist}.

\begin{figure}
\includegraphics[scale=0.5]{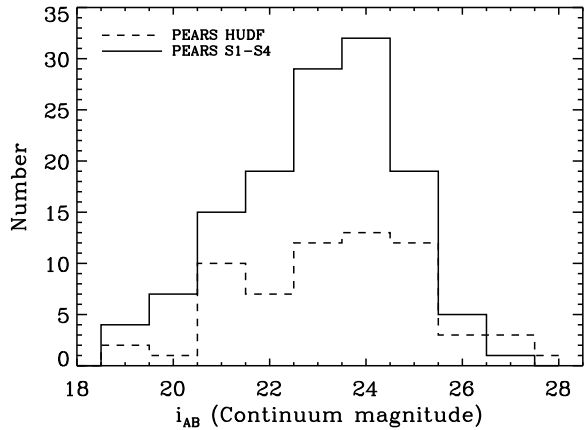}
\caption{The distribution of ELG continuum magnitudes peaks around $i_{AB}' \sim 24$ mag for both the HUDF and the PEARS South Fields 1---4 data.  The HUDF distribution is somewhat more uniform, owing in part to a larger fraction of faint objects due to its greater depth.  The clumpy face--on spirals generally make up the bright end of the magnitude distribution, while many of the HUDF sources comprise most of the faint end.}
\label{fig:magdist}
\end{figure}

\begin{figure}
\includegraphics[scale=0.5]{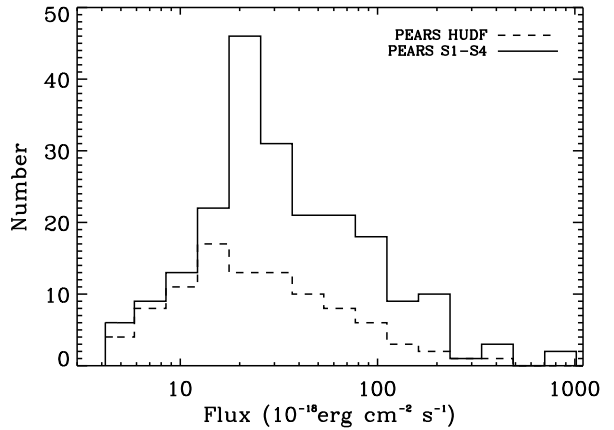}
\caption{The distribution of ELG emission--line fluxes peaks at $\sim$2.0\fluxsv\ for the PEARS South Fields 1--4 (20 HST orbits per field) and at $\sim$1.3\fluxsv\ for the deeper (40 HST orbits) PEARS HUDF.  }
\label{fig:fluxdist}
\end{figure}

\begin{figure}
\includegraphics[scale=0.5]{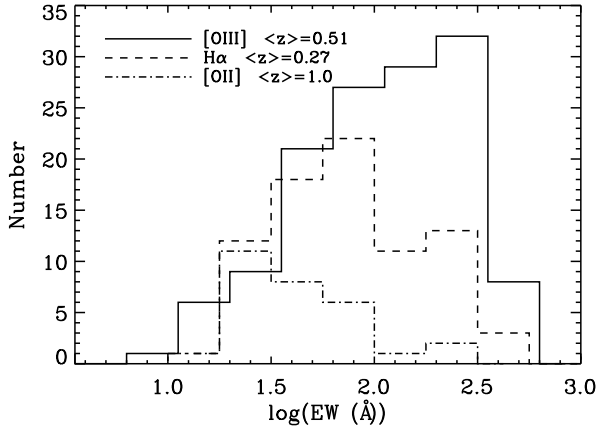}
\caption{The distribution of rest--frame equivalent widths of the three most common emission lines in our sample.  The median equivalent widths are 119\AA, 73\AA, and 36\AA\ for \OIII, \Ha, and \OII\ respectively.  The average redshifts of the three species is shown. }
\label{fig:ewdist}
\end{figure}

\begin{figure}
\includegraphics[scale=0.6]{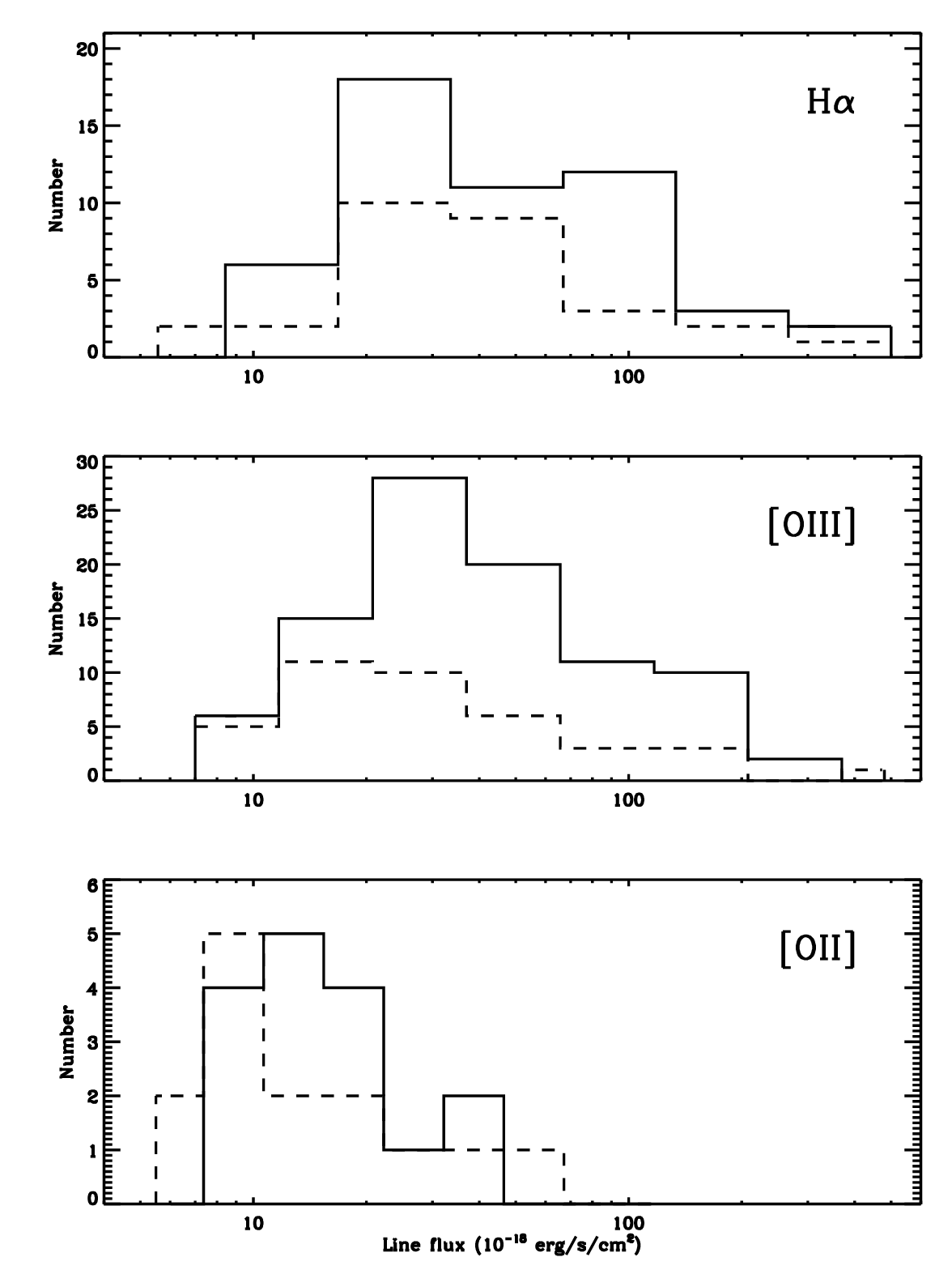}
\caption{The emission--line flux distributions peak at $\sim$2.5\fluxsv, $\sim$3.0\fluxsv, and $\sim$1.3\fluxsv\ for \Ha, \OIII, and \OII\ respectively for the 20--orbit/field PEARS data (four fields).  The PEARS HUDF line fluxes peak at slightly fainter values ($\sim$1.1\fluxsv\ for \OIII\ and $\sim$9.0\fluxei\ for \OII).}
\label{fig:fluxdistmulti}
\end{figure}

Figure~\ref{fig:magdist} shows the $i'_{AB}$-band continuum magnitude
distribution of the 192 ELGs in the PEARS South fields.  The
distribution peaks around $i'_{AB}=$24 mag for both the HUDF and the PEARS
South Fields 1--4, although the falloff at fainter magnitudes is more
pronounced for the shallower South Fields 1--4 data.  The 2D method
described here is optimized to find distinct emitting knots that often
are present in relatively bright galaxies---for example,
face--on spirals with large star--forming regions.  These generally
make up the bright-end of the magnitude distribution shown here.  The
fainter tail of the magnitude distribution is comprised largely of
objects from the deeper HUDF pointing.  The distribution of
emission--line fluxes for all 320 emission lines, regardless of
species, is shown in Figure~\ref{fig:fluxdist}.
Figure~\ref{fig:fluxdistmulti} shows that distribution for each of the
three most common emission lines in our sample: \Ha, \OII, and \OIII.
The flux distribution for the sample peaks at $\sim$1.9 \fluxsv\ for the
20--orbit/field PEARS data (four fields) and falls off at lower values
due to incompleteness of the data (Figure 5).  The peak is at a slightly
fainter flux for the deeper PEARS HUDF at $\sim$1.2 \fluxsv.

\begin{figure}
\includegraphics[scale=0.5]{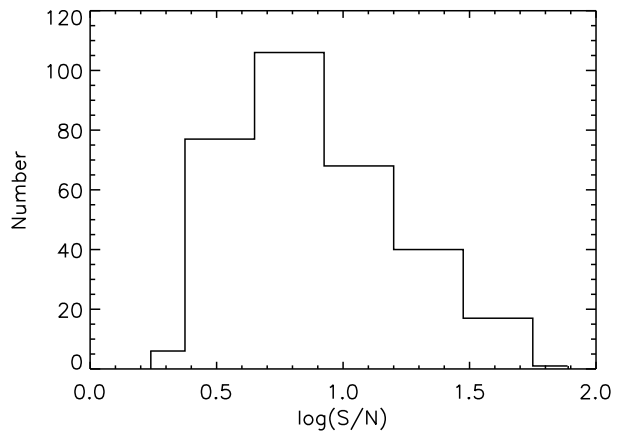}
\caption{Distribution of signal--to--noise for all derived line fluxes.  The average S/N for the sample is 11.8.  This average increases to S/N$=$12.3 when the weaker, blended \Hb\ lines are excluded.  Our detection method requires a relatively high S/N because the initial grism detection images are smoothed before source extraction is performed.  This is the reason we miss, e.g., lower S/N Ly$\alpha$ emitters (Rhoads \etal 2009). }
\label{fig:snhist}
\end{figure}

Given the ACS grism resolution, contamination of the dominant lines by
other nearby, unresolved lines is almost certainly present.  For
example, the \Ha\ line flux measurements will contain some
contribution from the [NII] $\lambda$$\lambda$6548,6584 lines.  The
magnitude of this contamination will differ for different galaxies, as
it depends on effective temperature, ionization, and metallicity.
Helmboldt \etal (2004) derive an [NII] correction as a function of $R$ band luminosity using the Nearby Field Galaxy Sample of spiral and irregular galaxies (Jansen \etal 2000).  Other grism surveys of ELGs have used
global corrections by Gallego \etal (1997), which also was derived
based on a local galaxy sample.  Our detection method serves to
produce individual galaxy knots in a wide array of morphological types
(as described in Section 3), and thus a global adoption of any one
[NII] contamination correction is not straightforward.  Therefore, the
measured \Ha\ fluxes are likely overestimates due to this
contamination but we do not adopt a global correction.  The amount of
contamination can range from a few percent for, e.g., blue compact
dwarf galaxies, which have unusually high ionization and low metallicity, to
the factors of 0.3 and 0.5 assigned by Gallego \etal (1997) and
Kennicutt (1992) respectively (however, the latter being for massive,
metal--rich galaxies).  For the Nearby Field Galaxy Survey (Jansen
\etal 2000), [NII]/\Ha\ ranges between 0.03--0.5 with a mean value of
0.27.  The signal-to-noise (S/N) distribution of the emission line fluxes is shown
in Figure~\ref{fig:snhist}.  The average S/N for the sample is 11.8.
This increases to 12.6 when the generally weaker, blended
\Hb\ line measurements are excluded.  Our detection methods outlined
above serve to produce a final sample of high-confidence detections.

The presence of dust affects our measurements, and thus the calculations of, e.g., the star--formation rate (Section 4.4.3) should be considered lower limits because no extinction correction was applied.  The \Hb\ flux in principle allows an estimation of extinction for the cases in which both \Hb\ and \Ha\ fall into the wavelength range of the grism and including \Hb\ results in a quanitatively better fit.  This is only possible for a very small percentage of objects and thus we do not apply a global correction based on only these few sources.  Both \Ha\ and \Hb\ are measured in the spectra of objects 38750, 40816, 75753, 78582, and 123859.  However, objects 40816 and 78582 are both X--ray sources and therefore likely AGN candidates (see Section 4.3).  Because of this, the emission line fluxes of these two sources are likely affected by the potential AGN component.  Using only the Balmer decrement and the Milky Way or LMC extinction law from Seaton (1979)---e.g., Calzetti \etal 1994, who find an average E(B-V) of 0.4 for starburst galaxies---gives E(B-V) values of 0.60, 0.26, and 0.50 for objects 38750, 75753, and 123869 (e.g., those not X--ray detected) respectively.

\subsection{Grism Redshifts}
\begin{figure}
\includegraphics[scale=0.5]{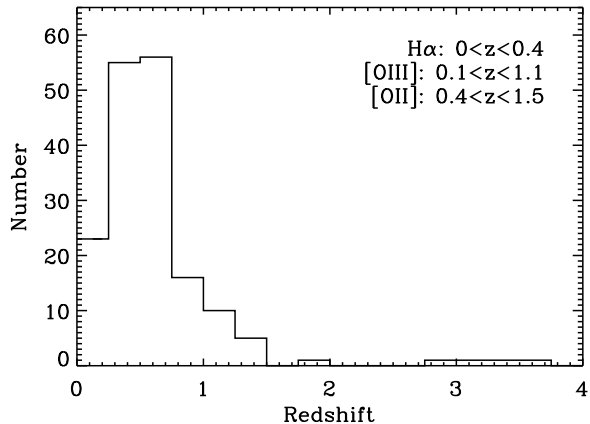}
\caption{Emission--line galaxy redshift distribution.  The G800L grism is sensitive from 6000-9500 \AA, which yields the most common emssion lines---\Ha, \OIII, and \OII\ in the wavelength ranges of z=0--0.4, 0.1--1.1, and 0.4--1.5 respectively.  The \OIII\ line is the most common, and thus the peak is near z$\sim$0.5.  The higher redshift objects are the more rare \CIII, \CIV, and MgII emitters.}
\label{fig:zdist}
\end{figure}

Of the 192 emission--line galaxies, 118 have new grism spectroscopic
redshifts based on our line identifications.  We find 8 galaxies
(Table 1) that previously had no reported redshift and that have two
lines, allowing determination of a grism redshift from the wavelength
ratios.  The redshift distribution of the sample is given in
Figure~\ref{fig:zdist}.  The redshift distribution peaks at z$\sim$0.5
and is determined by the most common emission lines within the grism
bandpass: \OIII, \Ha, and \OII.  This explains the lower redshift peak
compared to the peak in the general field galaxy redshift distribution.  The few
high--redshift objects in this plot are the more rare \CIII, \CIV, and
MgII line emitters.  All of these high--redshift sources in the CDF--S are
detected in the X--ray observations, and are thus likely AGN (Grogin
\etal 2009, in preparation).  The CDF-S X-ray sources are noted in Table 1.

\vfill\eject
\begin{figure}
\centering
\subfigure{
\includegraphics[scale=0.5]{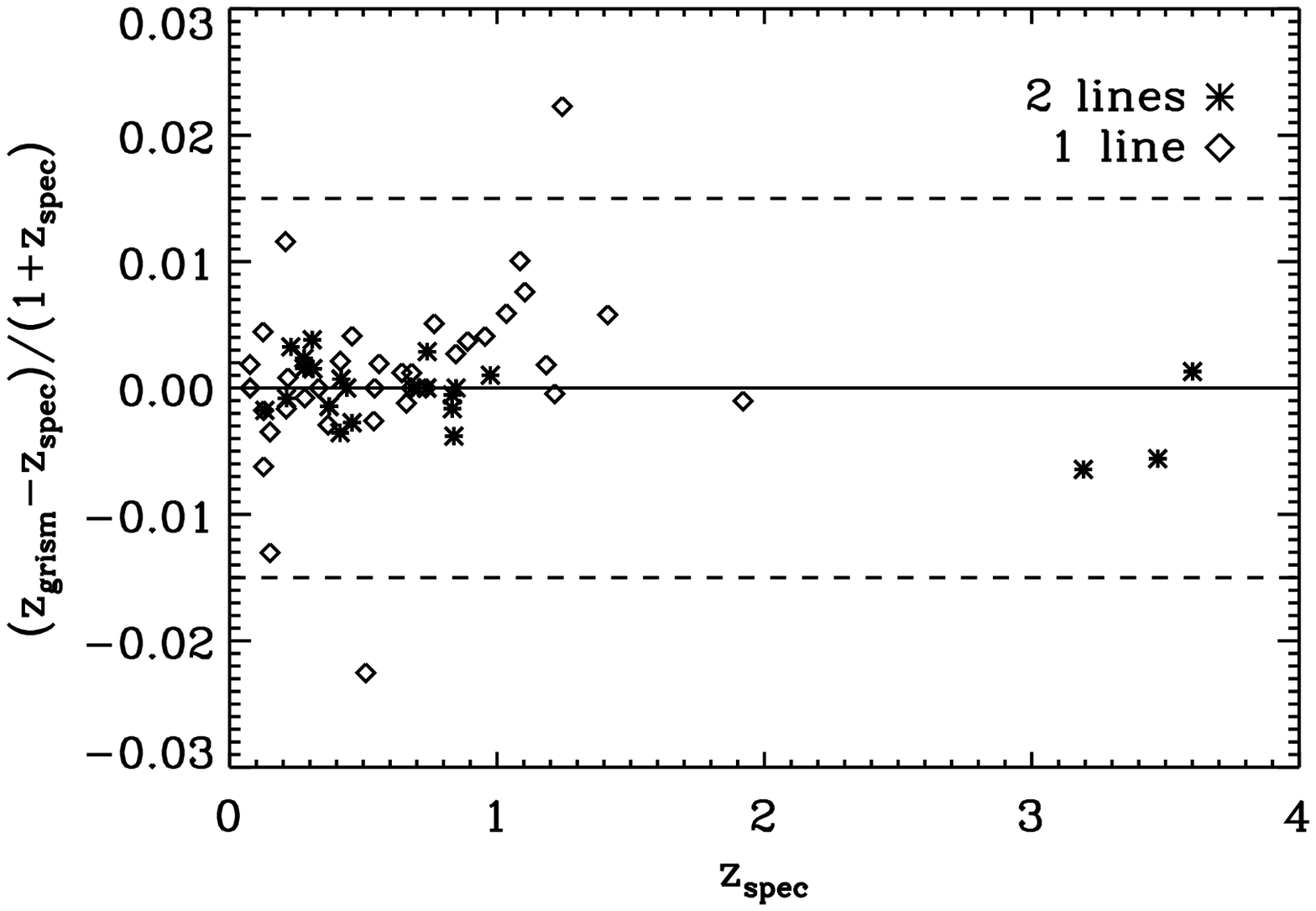}}
\hspace{-0.20cm}
\subfigure{
\includegraphics[scale=0.5]{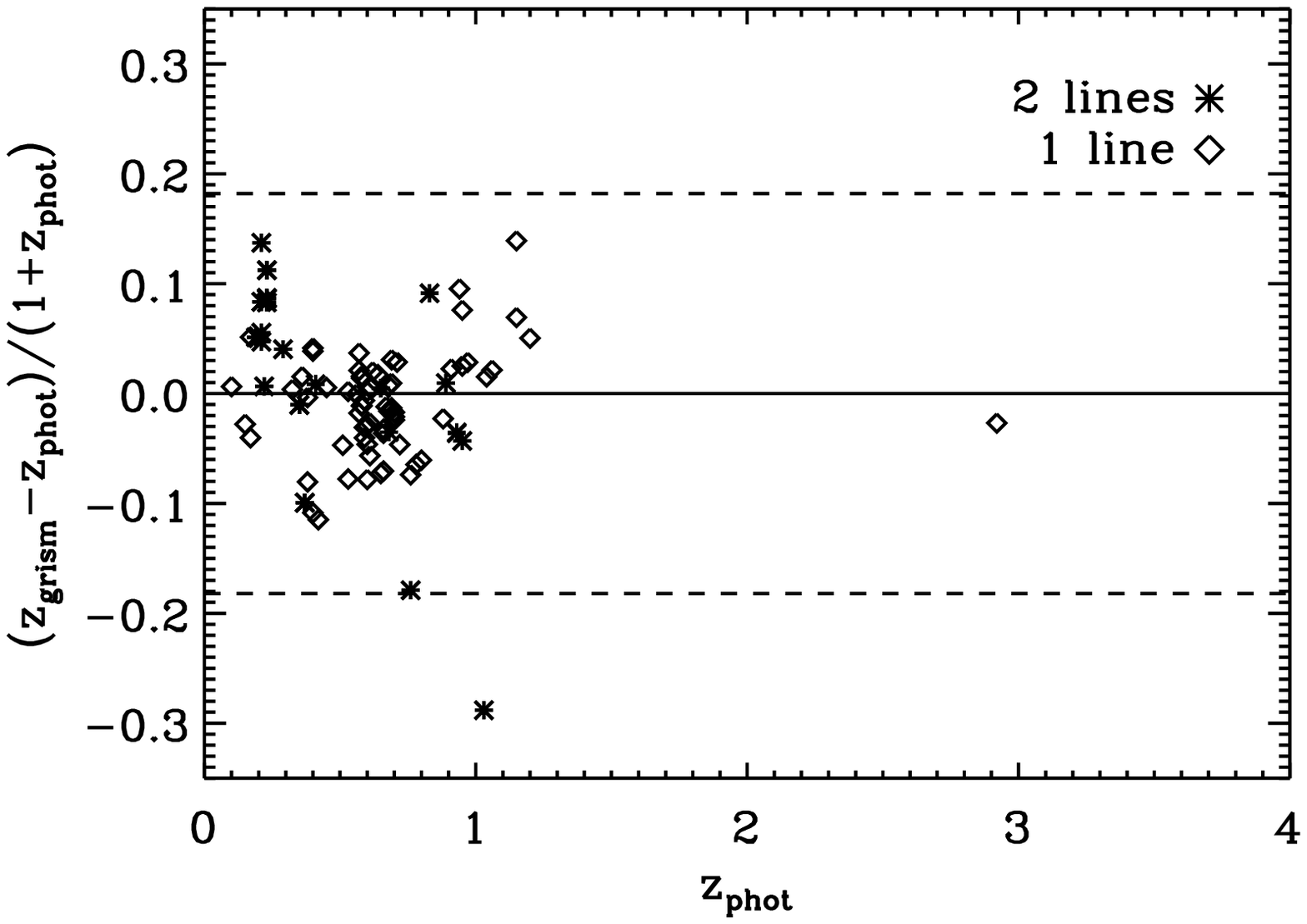}}
\caption{Comparison of available spectroscopic (top panel) and photometric (bottom panel) redshifts to the PEARS grism redshifts measured in this study, with 3$\sigma$ (dashed) lines shown.  31\% and 81\% of PEARS-South ELGs have previously--measured spectroscopic and photometric redshifts, respectively.  Comparison of grism to spectroscopic redshifts essentially serves to demonstrate the wavelength$/$redshift calibration accuracy of the PEARS grism data.  See section 4 for a discussion on outliers.}
\label{fig:zcomp_spec}
\end{figure}
\vfill\eject

In Figure~\ref{fig:zcomp_spec} we show
comparisons of our calculated grism redshifts to the available
photometric and spectroscopic redshifts for the ELGs.  As mentioned,
for objects with only a single emission line, any previously available redshift
was used to initially identify the line.  This was accomplished in the
cases where the line wavelength falls within the expected wavelength
based on that object's previously--measured redshift, within the redshift (and inherent
grism) errors.  

\begin{figure}
\includegraphics[scale=0.5]{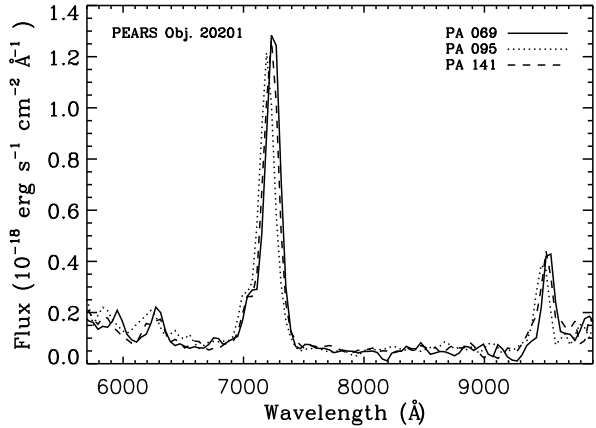}
\caption{An example spectrum from PEARS Object 20201 at a redshift of z=0.445, exhibiting the blending of \Hb and \OIII.  \Ha is also visible near the red end of the spectrum.  An \Hb\ ``bump'' is clearly seen near 7000\AA, though not resolved from the stronger \OIII\ blend.  In total, 31 galaxy spectra had better $\chi^{2}$ fits when the \Hb\ line was included.  H$\gamma$ and marginal H$\delta$ are detected here, near 6300\AA\ and 6000\AA\ respectively.}
\label{fig:exspec}
\end{figure}

Comparison of grism--spectroscopic redshifts computed here to previously--existing spectroscopic redshifts serves
to demonstrate the wavelength accuracy of the grism, which is shown in
Figure~\ref{fig:zcomp_spec}.  The dispersion about the mean is 0.005
and two objects are \cge 3$\sigma$ outliers: PEARS Objects 72509 and
17362, both of which are single--line detections with relatively low
$S/N<3$ and likely represent wavelength calibration issues.  As
expected, the dispersion about the mean in the photometric/grism
redshifts is greater at 0.06, with the greatest $\Delta$z=0.585
(PEARS Object 52502, the only 3$\sigma$ outlier).  This object has two
emission lines with S/N$>$5, providing a secure grism redshift
based on the wavelength ratio. Object 20201, which was only marginally
within 3$\sigma$ of the photometric redshift also has two high S/N
emission lines, as well as a clear \Hb\ ``bump'' in the \OIII\ line
profile, further confirming its identification
(Figure~\ref{fig:exspec}).  Thus for these two outlying objects, we are
confident that the grism redshift calculated here is correct.

\subsection{Line Luminosities \& Star--formation Rates of the ELGs}
Table 1 lists the line luminosities for the objects in our sample.
The median \Ha\ line luminosity is 8.3$\times$$10^{39}$ ergs s$^{-1}$, and the lowest luminosity is 2.5$\times$$10^{38}$ ergs s$^{-1}$.  As a
comparison, Drozdovsky \etal (2005) find a median \Ha\ line luminosity
of 2.7$\times$$10^{40}$ ergs s$^{-1}$ from the ACS Grism Parallel
Survey.  The typical local $L^{*}$(\Ha)=7.1$\times$$10^{41}$ ergs
s$^{-1}$ (Gallego \etal 1995) and $L^{*}$(\Ha)=3.6$\times$$10^{42}$
ergs s$^{-1}$ at z=1.3 (Yan \etal 1999).  The median \OIII\ and \OII\ line
luminosities are 2.8$\times$$10^{40}$ ergs s$^{-1}$ and 6.7$\times$$10^{40}$ ergs s$^{-1}$ respectively.
About 96\% of our emitting regions have luminosities \emph{L}\cge$10^{39}$ ergs s$^{-1}$.

\begin{figure}
\includegraphics[scale=0.5]{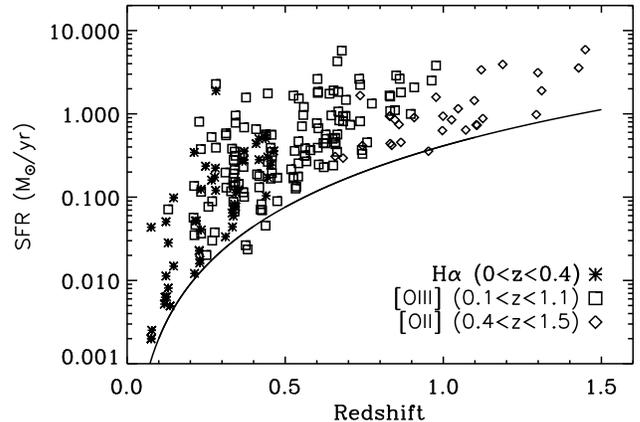}
\caption{Star formation rates as a function of redshift based on the line luminosities of the ELGs.  We see the expected bias of higher SFRs at higher redshifts, due to the detection limits.  These SFRs are uncorrected for extinction and are thus lower limits.  The approximate empirical detection limit---derived from the average limiting flux of all three lines---is shown for the (deepest) PEARS HUDF data.}
\label{fig:sfr}
\end{figure}

We present the star-formation rate (SFR) as a function of redshift of
our ELG sample in Figure~\ref{fig:sfr}.  SFRs are calculated using the
calibrations of Kennicutt (1998) for \Ha\ and \OII:

\begin{equation}
$$SFR$_{\Ha}$ (\Mo yr$^{-1}$)=7.9$\times$$10^{-42}$ L(\Ha) ({\rm erg\ s$^{-1}$})$$
\end{equation}

\begin{equation}
$$SFR$_{[OII]}$ (\Mo yr$^{-1}$)=1.4$\times$$10^{-41}$ L([OII]) ({\rm erg\ s$^{-1}$})$$
\end{equation}

respectively for solar abundances and a Salpeter IMF for 0.1-100\Mo.  The
\Ha\ luminosity is a direct measure of the ionizing output of a
stellar population (under case--B recombination) and thus can be
related directly to the massive star--formation rate.  In particular,
it probes the formation of the ionizing O stars, and thus is the most
secure line in determining the SFRs.  The SFR based on \OII\ line
luminosity is less secure, as differences in metallicity and other
local environmental properties play a larger role in the oxygen lines
(Kewley \etal 2001; Jansen \etal 2001; Kewley \etal 2004).  Kennicutt (1998), for example, reports a $\sim$30\% uncertainty on the \OII\ SFR calibration.  However,
the \OII\ line is still calibrated well enough to deduce SFRs for
galaxies at higher redshift (Cowie \etal 1996, Kennicutt 1992,
Gallagher \etal 1989).  We use the Kennicutt (1998) calibrations for
the \Ha\ and \OII\ emitters in the PEARS-South ELG sample presented
here.

The determination of SFRs from \OIII\ line luminosities is not as straightforward, since the \OIII\ flux depends quite strongly on metallicity and gas temperature (Kennicutt \etal 2000, Kennicutt 1992), and SFRs derived from the \OIII$\lambda$5007 line have a typical scatter of 3--4 when uncorrected for reddening (Moustakas \etal 2006).  However, the \OIII\ line has been used to gain crude lower limits on the SFR (Maschietto \etal 2008; see also Teplitz \etal 2000 for a discussion of \OIII\ SFRs for LBGs).  Maschietto \etal (2008) arrive at a lower limit of SFR$_{[OIII] (5007)}$($\Mo$yr$^{-1})<$3.3 $\times$ 10$^{-42}$ L$(\OIII)$ ergs s$^{-1}$ for their sample of 13 star--forming galaxies.  With the ACS G800L grism resolution, the \OIII $\lambda$$\lambda$4959, 5007 doublet and \Hb\ are blended, and while our fitting technique does fit the blended \OIII $+$ \Hb\ feature, some cross--contamination of the lines is likely.  Many of the galaxy knots that contain \OIII\ emission originating from star--formation (and not from AGN as described in Sec. 4.3) also have either \Ha\ or \OII\ lines in their spectra, so in these cases, it is clearly best to use the more direct \Ha-- or \OII--deduced SFR.  For the emitting regions in which only an \OIII\ line is detected---due to the \Ha\ or \OII\ lines falling out of the grism bandpass---we derive the \OIII\ SFR by using the \OIII\/:\Ha\ ratio from the galaxy knots that do have both emitting lines.  Since the \OIII\ $\lambda$$\lambda$4959, 5007 doublet is blended, we use 0.66$\times$L$_{\OIII}$ in order to estimate the contribution from the $\lambda$5007 line only.  We thus arrive at:

\begin{equation}
$$SFR$_{[OIII]}$ (\Mo yr$^{-1}$)=(6.4 $\pm$ 4.0) $\times$$10^{-42}$ L([OIII]) ({\rm erg\ s$^{-1}$})$$
\end{equation}


While there is large scatter in the \OIII--derived SFR, we find no indication of nonlinearity in the relation of \Ha\ and \OIII\ for this subsample of ELGs.  The possible presence of residual blended \Hb\ flux 
described above provides an additional source of error to the \OIII\ flux derivation.
However, in all cases, we did not apply extinction corrections, and
thus the implied SFRs presented here are in general lower limits.
In addition, we assume that most of the galaxies' active
star formation is occurring in these emission--line regions, but note that
the sample is incomplete in the sense that only the brightest knots of
the galaxies are detected, and diffuse emission is missed in our
method.  Figure~\ref{fig:sfr} shows the expected bias of lower SFRs
at lower redshift.  This in general follows calculations performed in
similar studies, e.g., Drozdovsky \etal 2005, who computed SFRs of a grism--selected sample of ELGs.  

\subsection{Potential AGN Candidates Among the ELG Sample}
\begin{figure}[htbp]
\includegraphics[scale=0.5]{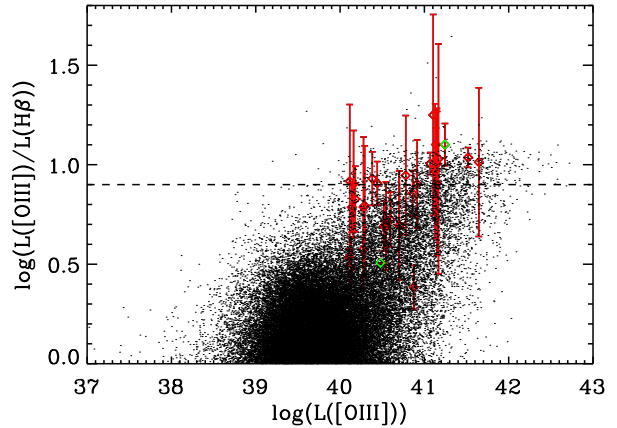}
\caption[Estimate of emission--line sources' excitation as a function of SFR.]
{\OIII\ to \Hb\ flux ratios of PEARS ELGs (red diamonds) compared to those from the SDSS AGN catalog (dots; Kauffmann \etal 2003).  As discussed in Sec. 4, the PEARS objects with \OIII:\Hb\ \cge 8 (dashed line) are probable AGN.  The two green diamonds are the two X--ray confirmed objects that have both \OIII\ and \Hb\ measured (PEARS Objects 40816 and 78582).}
\label{fig:sfroiihb}
\end{figure}

Adjusting our line fitting algorithm to include \Hb\ fits allows us to
gain a crude estimate of excitation.  In Figure~\ref{fig:sfroiihb}, we
show the \OIII:\Hb\ line ratio compared to a large sample of SDSS AGN.
Kauffmann \etal (2003) compare this line ratio to
[NII]$\lambda$6583$/$\Ha\ and thus define a region of likely AGN (as
compared to starburst galaxies) in a BPT diagram (Baldwin, Phillips, \& Terlevich 1981;
see also Kewley \etal 2001).  In the grism data, the [NII] line is
blended with \Ha\ and is not possible to deblend, as is the case with
some objects for \OIII\ and \Hb, and thus a BPT diagram is not
possible to construct from the PEARS ELGs.  However, starburst galaxies with
\OIII:\Hb$\ge$8---taking into account the blending of the
\OIII\ doublet---are extremely rare based on the starburst/AGN
demarcations made by both Kauffmann \etal (2003) and Kewley \etal
(2001), which include effects of metallicity and dust.  We thus
conclude that the PEARS objects that lie above this threshhold are
potential AGN candidates among our ELG sample.  There are a total of
27 ELGs that have both \OIII\ and \Hb\ measured in at least one of
their knots.  Of these, 3 have F(\OIII)/F(\Hb)$\ge$8 within their
observational uncertainties, and another 14 have likely
F(\OIII)/F(\Hb)$\ge$8 but with larger errors (see Figure~\ref{fig:sfroiihb}).

In addition to high \OIII:\Hb\ line ratios, high X--ray luminosities
are also strong indicators of AGN activity.  Grogin \etal (2007) and
Grogin \etal (2009, in preparation) investigate CDF X--ray sources
that fall within the PEARS area.  In total, 17 of the emission--line
sources detected in this study overlap with the Grogin \etal PEARS X--ray
sample.  Of these 17, 8 objects have X--ray luminosities $L_{X}$$\ge$$10^{42}$
ergs s$^{-1}$ and are thus likely AGN.  These $L_{X}$$\ge$$10^{42}$ ergs
s$^{-1}$ sources display mainly the expected AGN lines (e.g., \CIII,
\CIV, and [MgII]).  All matches to CDF-S X-ray sources are noted in
Table 1.  

Of the PEARS emission--line sources with both the \OIII\ and \Hb\ lines measured, two are also X--ray
sources (Grogin \etal 2009, in preparation), but with $L_{X}$ $<$ $10^{42}$ ergs s$^{-1}$.  One of these
two objects, PEARS Object 40816 at redshift z=0.281, has a quite
high flux ratio F(\OIII)$/$F(\Hb)=12.6 and an X--ray luminosity of
$L_{X}$=$2.1\times10^{41}$ ergs s$^{-1}$.  Object 40816's line
emission originates from the galaxy's nucleus.  The galaxy appears to
be interacting with a nearby disk galaxy (PEARS Object 35818) with a
tidal stream in between the two objects.  Given this PEARS Object
40816's high F(\OIII)$/$F(\Hb) and moderate $L_{X}$ values, one can
interpret this source as being a potential obscured
interaction--induced AGN.  The other object, PEARS Object 78582 with
redshift z=0.454, has a flux ratio F(\OIII)$/$F(\Hb)=3.2 and
$L_{X}$=$1.8\times10^{41}$ ergs s$^{-1}$.  This source appears
spheroidal with signs of tidal debris and/or interaction with PEARS
Object 78762.  Object 78582 is thus likely a regular star--forming
galaxy with starburst--related X--ray emission, given its lower
F(\OIII)$/$F(\Hb) value.  As Figure~\ref{fig:sfroiihb} demonstrates,
the PEARS AGN candidates based on the \OIII:\Hb\ ratio reside mainly
on the upper right locus of the SDSS sample (black dots).  The lack of objects with
lower excitation is likely a result of the de--blending of the
\OIII\ and \Hb\ lines---as was noted in Section 3.3, the \Hb\ line was weaker by a factor of at least 2 in spectra where both lines were fit.  We thus
conclude that inclusion of \Hb\ in the line fitting procedure when
possible provides a way in which to select probable AGN from the grism
data for follow--up study and confirmation.

\subsection{High--redshift Star--forming Regions}
\begin{figure}
\centering

\hspace{-0.2cm}
\subfigure{
\includegraphics[scale=0.127]{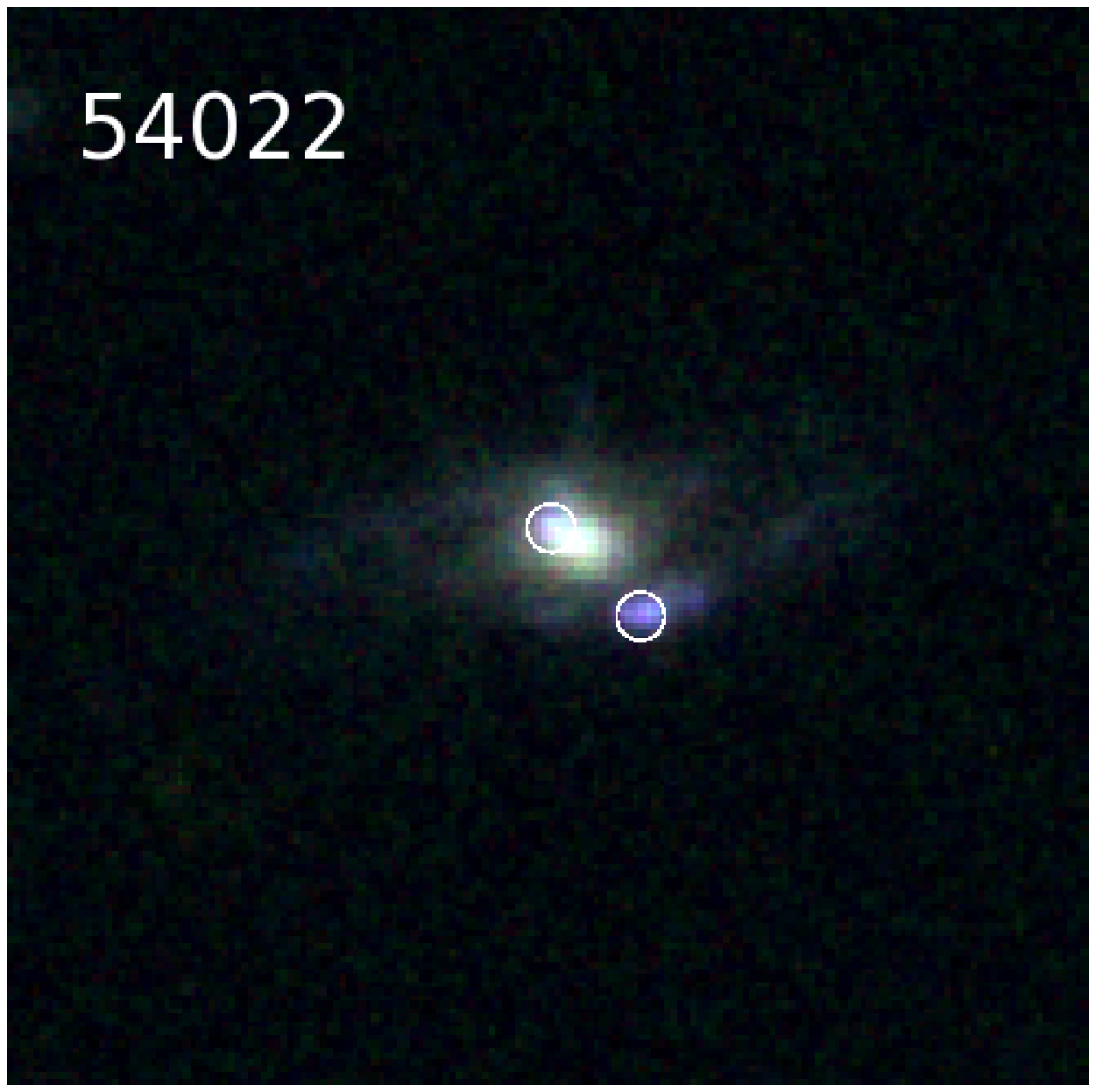}}
\hspace{-0.2cm}
\subfigure{
\includegraphics[scale=0.127]{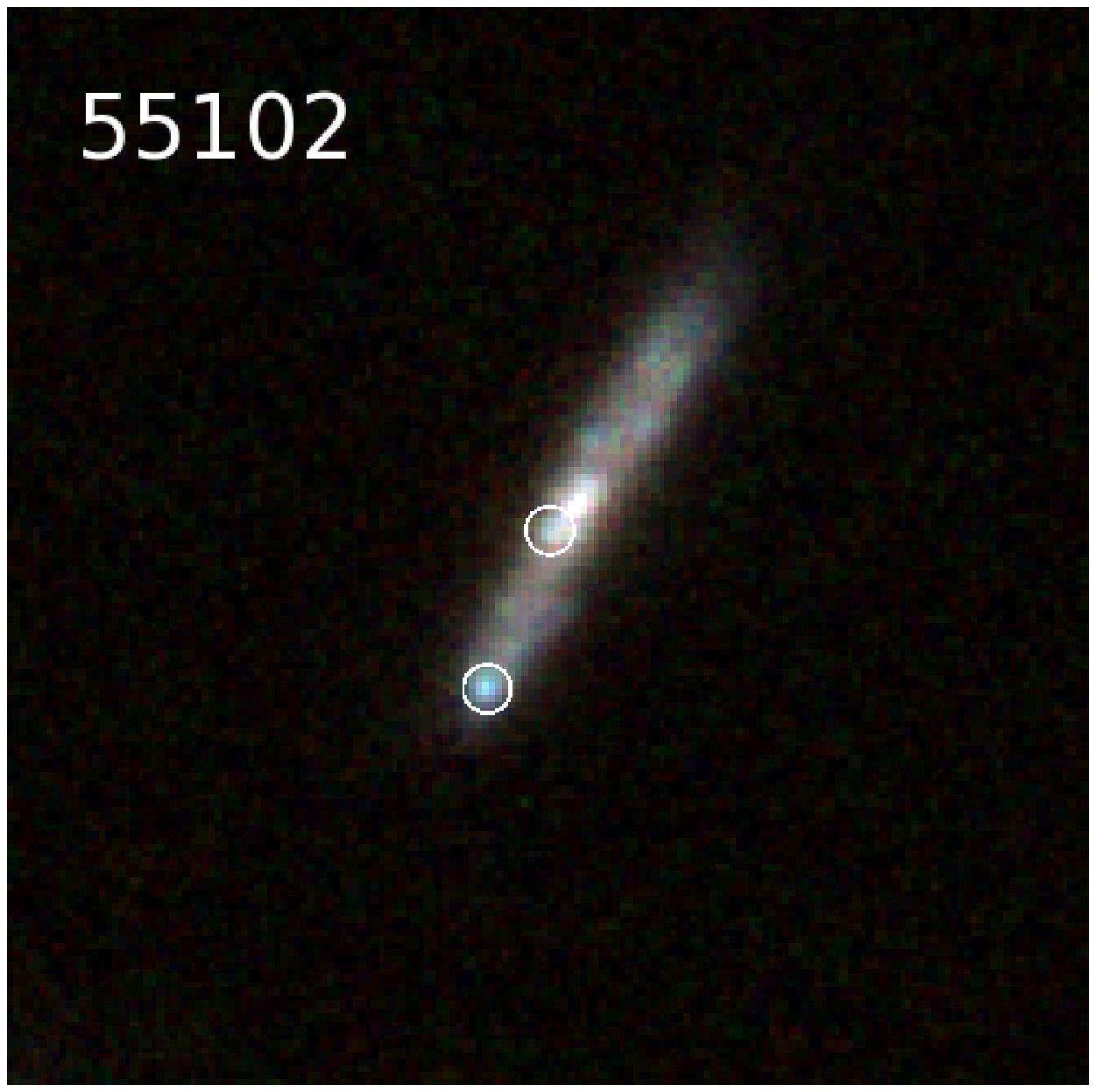}}
\hspace{-0.2cm}
\subfigure{
\includegraphics[scale=0.127]{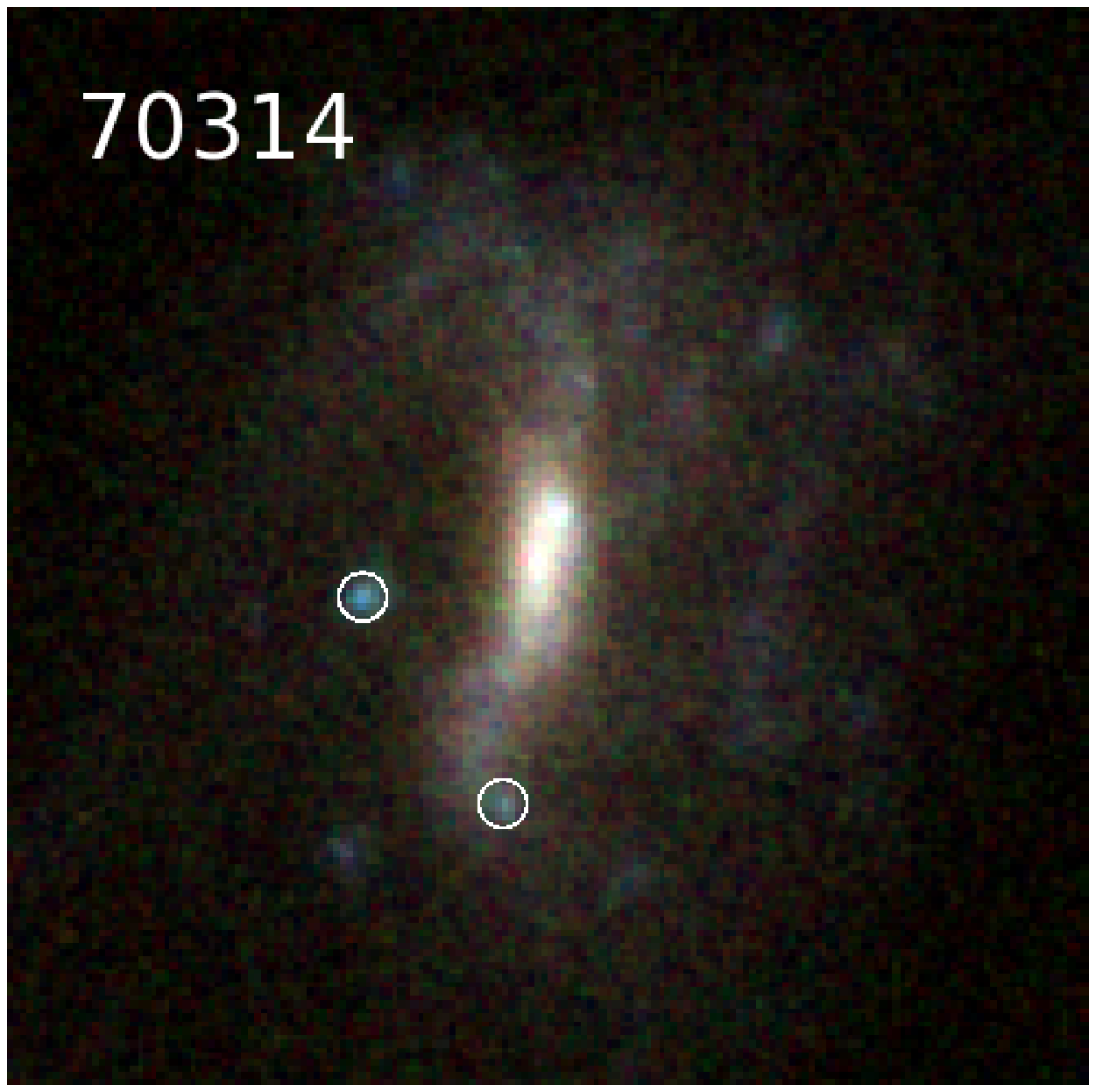}}
\hspace{-0.2cm}
\subfigure{
\includegraphics[scale=0.127]{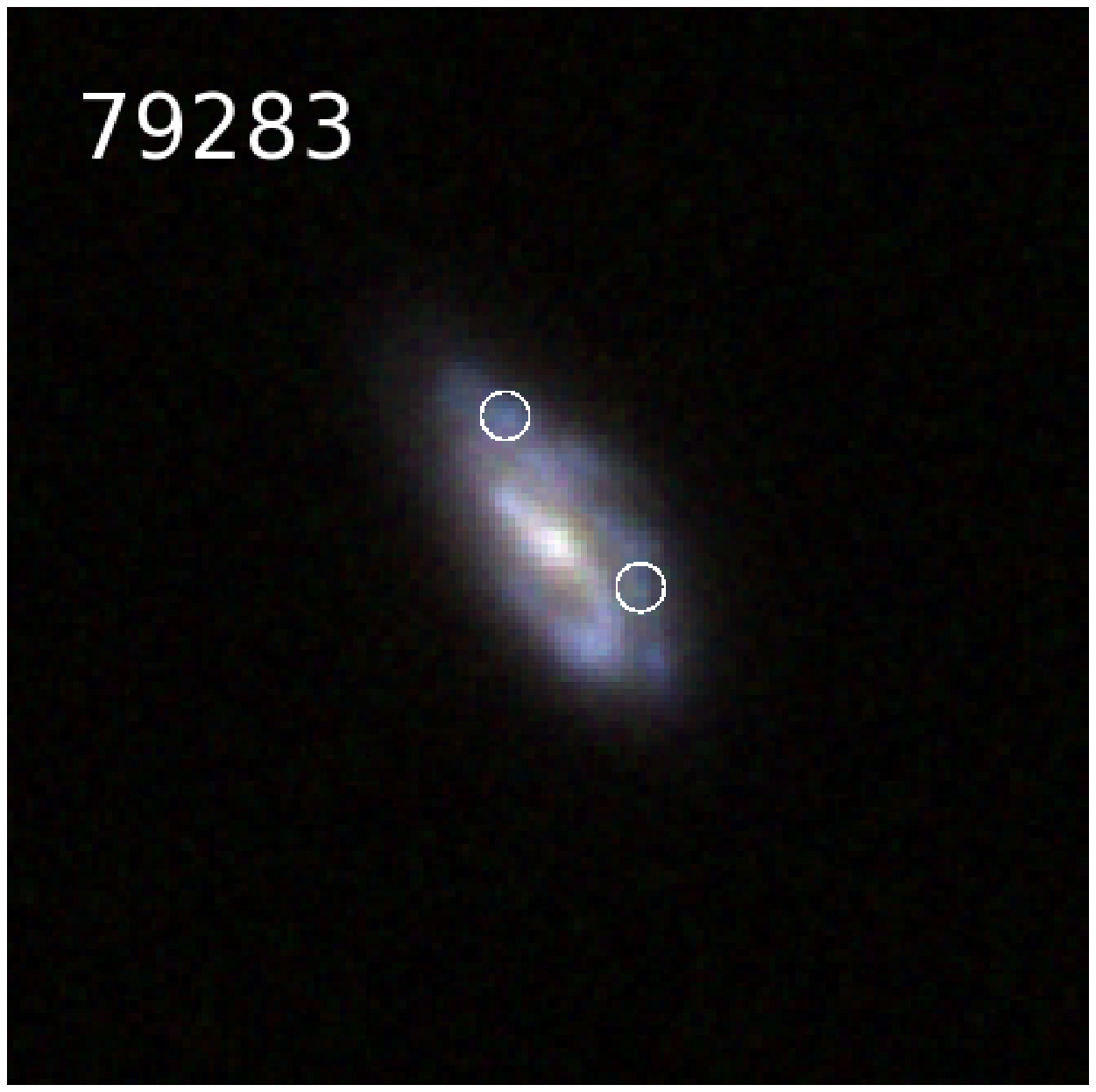}}
\hspace{-0.2cm}
\subfigure{
\includegraphics[scale=0.127]{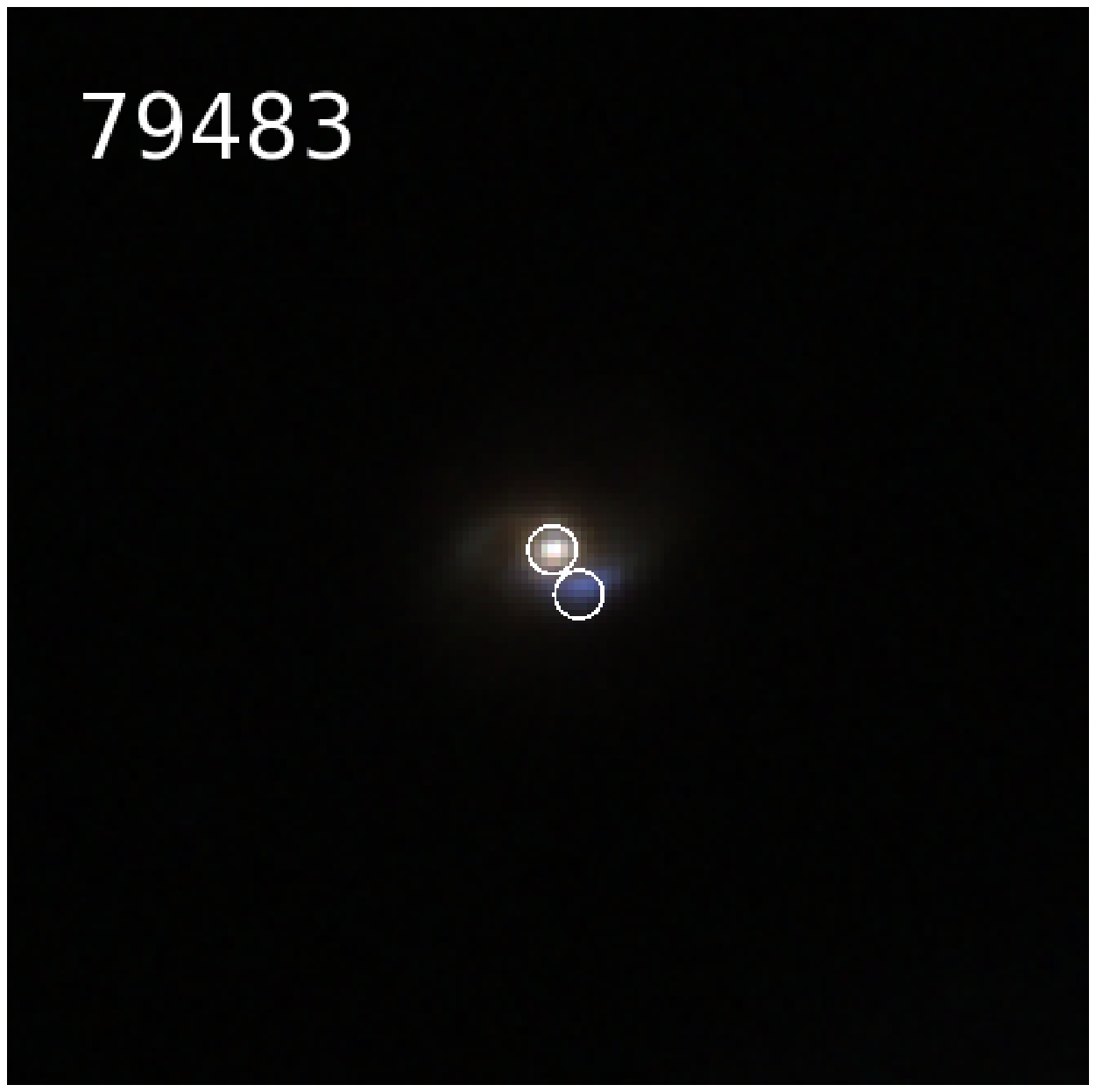}}
\hspace{-0.2cm}

\hspace{-0.2cm}
\subfigure{
\includegraphics[scale=0.127]{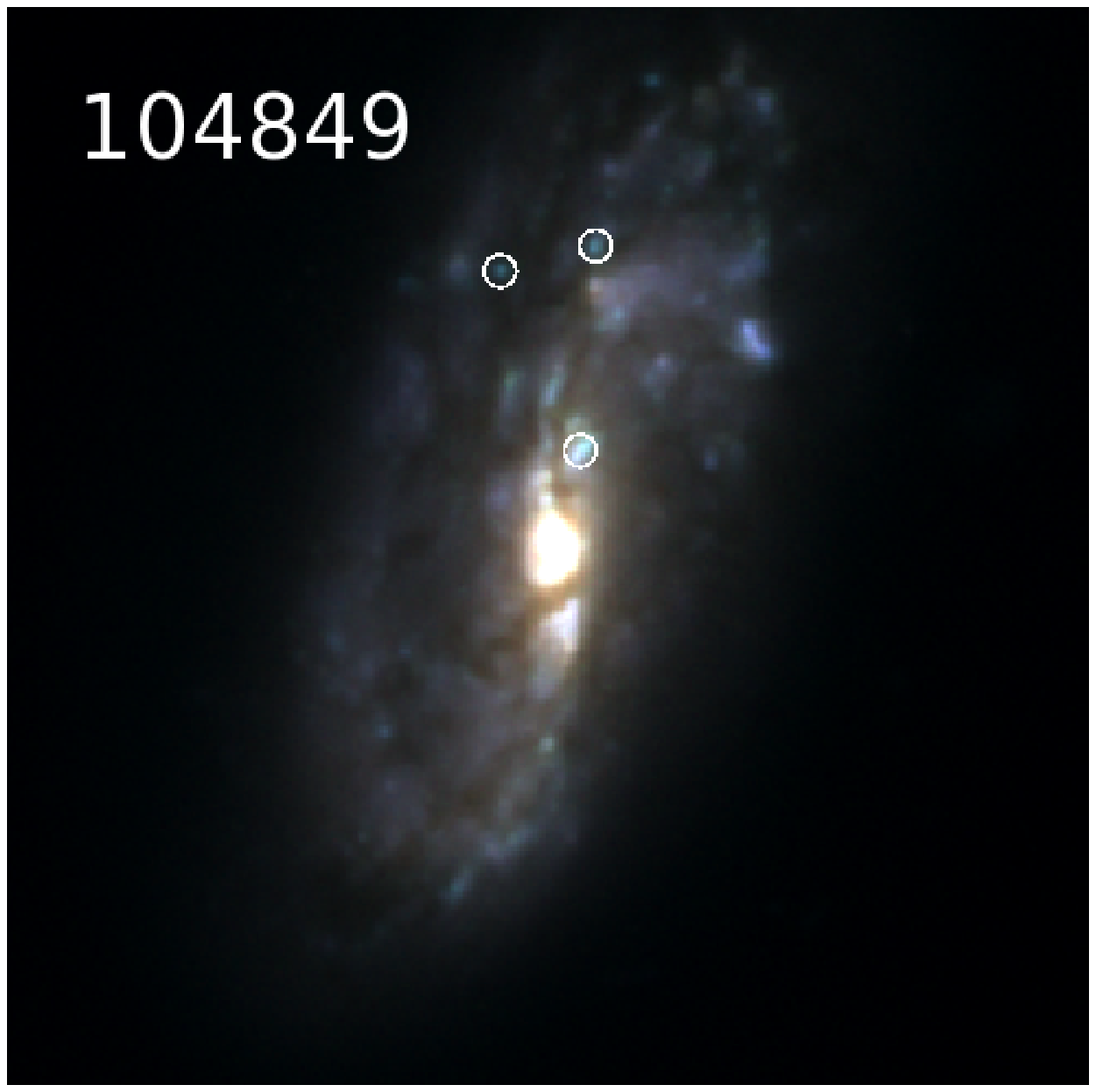}}
\hspace{-0.2cm}
\subfigure{
\includegraphics[scale=0.127]{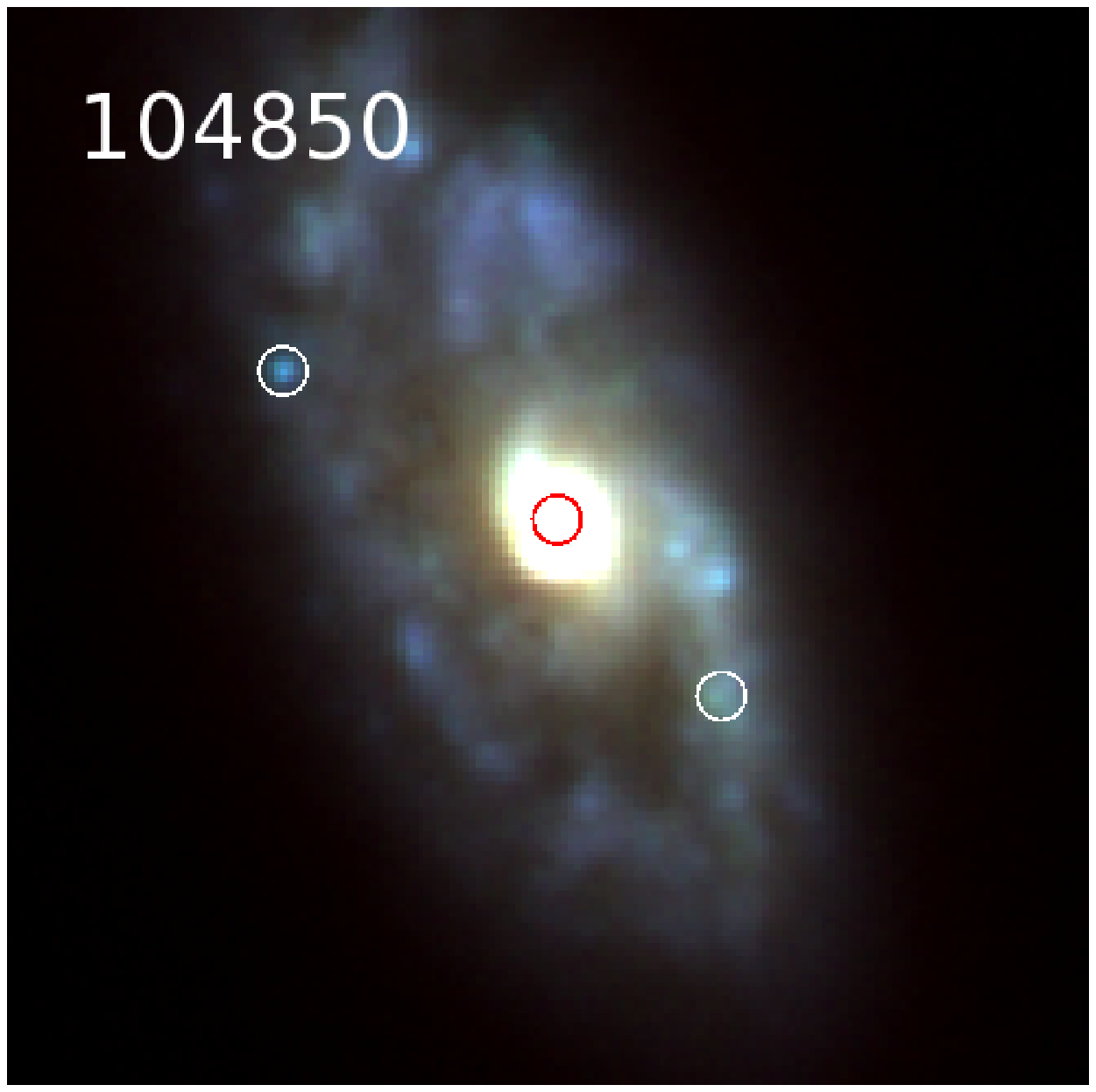}}
\hspace{-0.2cm}
\subfigure{
\includegraphics[scale=0.127]{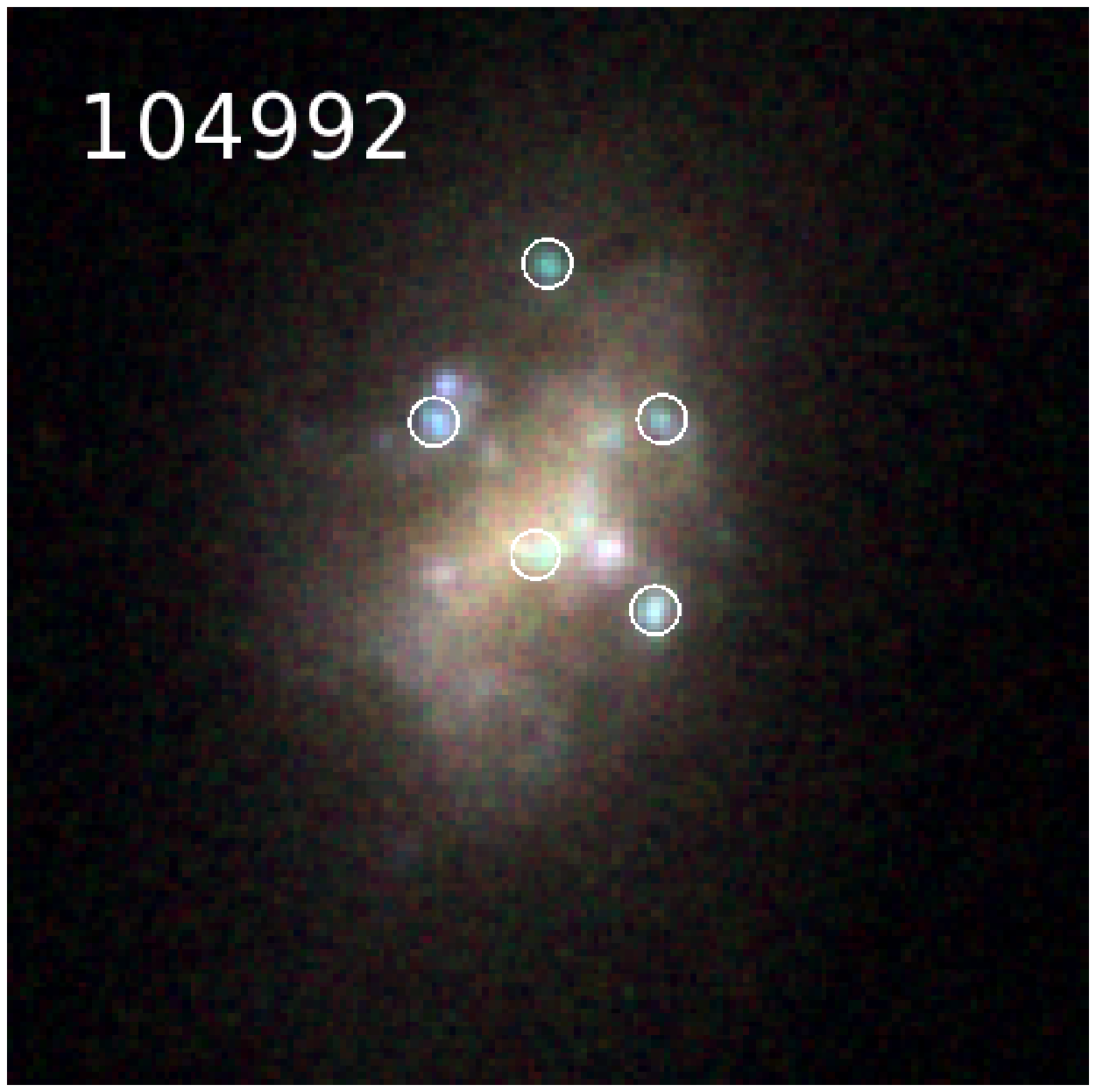}}
\hspace{-0.2cm}
\subfigure{
\includegraphics[scale=0.127]{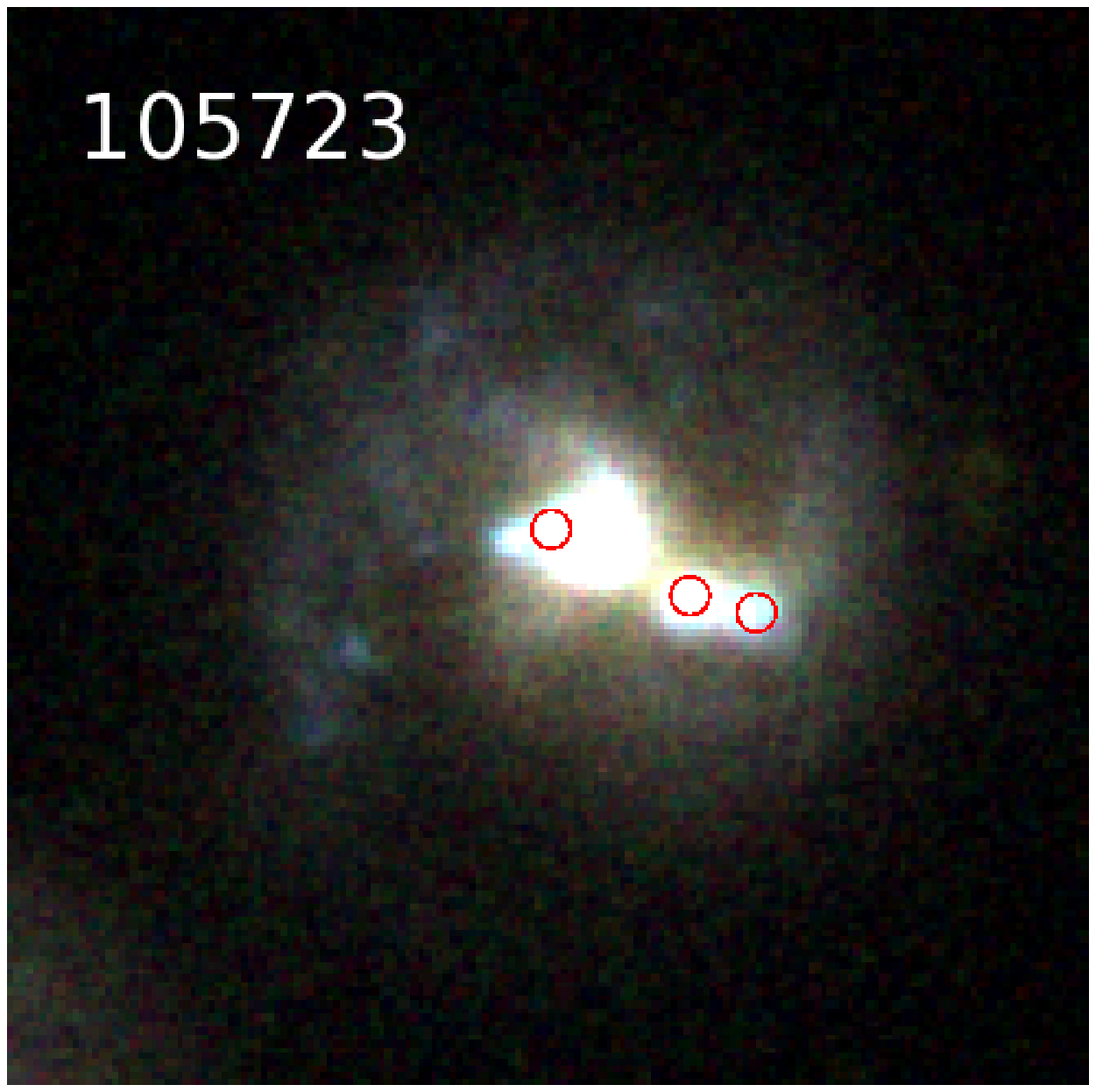}}
\hspace{-0.2cm}
\subfigure{
\includegraphics[scale=0.127]{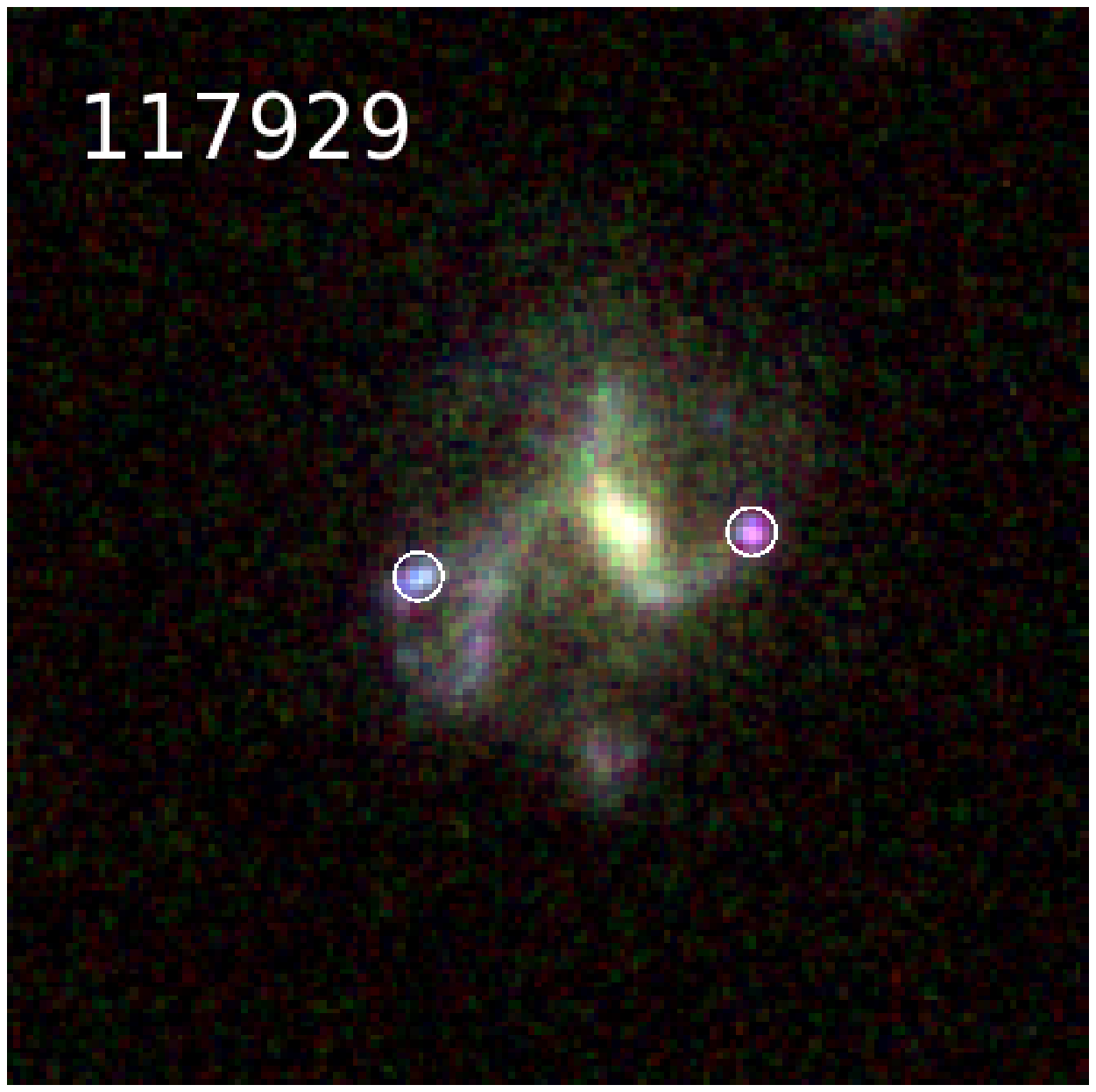}}
\hspace{-0.2cm}

\caption{Subset of PEARS ELGs with multiple emitting knots as described in Section 4.3, from which the radial distribution of star--forming knots is derived.  PEARS IDs are given in upper--left corners of stamps, which are 5 arcsec on a side.  Circles indicate region of line emission (colored circles are for visual aid only in bright regions).  The automated 2D--detection method is optimized to detect line emission in galaxy knots as shown here.  The radial distribution of the galaxy knots shown here is given in Figure~\ref{fig:radialdist}.}

\label{fig:multknotsmontage}
\end{figure}

One of the main advantages of the 2D--detection method used for this
study is the detection of emission lines in distinct star--forming
regions within galaxies at intermediate redshift---regions that would
not have been detected if the spectrum of the entire galaxy was
extracted (Figure~\ref{fig:compspec}).  In $\sim$12\% of galaxies we
find multiple emitting knots (Figure~\ref{fig:multknotsmontage}).
Many of these multiple--knot emitters are clumpy spirals with
distinct star--forming regions.  In total, 25 galaxies have multiple
emitting knots.  Within these galaxies, there are 59 such knots with
83 emission lines total---the majority of which are \Ha.  The median
redshift of the subsample of multiple--emitting knot ELGs is
z$=$0.336, and the highest redshift multiple--knot emitter is at
z$=$0.653.  While properties of local individual HII regions have been
studied for some time (e.g., Hodge 1969; Shields 1974; Shields 1990; McCall,
Rybski, \& Shields 1985; Zaritsky, Kennicutt, \& Huchra 1994; Gordon
et al. 2004; Kennicutt 1984), grism surveys such as PEARS---combined
with the 2D--detection method used here---are useful for finding
spectra of individual intermediate--redshift star--forming regions.
As discussed in Section 4.2, our detection limit serves to produce a
sample of mostly \emph{giant} star--forming regions, which have been
studied extensively in the local universe since they are sites of the
most extreme star formation known (e.g., Shields 1990;
Giannakopoulou-Creighton \etal 1999).  We find that within an individual galaxy, the \Ha--derived SFR typically differs by a factor of two or three between knots.  The most extreme differences in SFRs across
individual galaxies do not occur in the face--on spirals that are quite
common in the subsample of multiple--emitting knot galaxies, but in
clumpy galaxies with clear merger signatures.  This effect is not
unexpected, since mergers are known to induce enhanced star formation
activity which is revealed through the galaxies' emission lines.
Regions of the galaxy that are undergoing more intense physical
alterations due to the merging activity presumably exhibit more
intense star formation.

One of the questions that can be addressed through the 2D--detection technique
concerns how the galaxies' giant star--forming knots are distributed
radially within each galaxy.  As with giant HII regions generally,
these radial distribution studies have typically been performed on
nearby spiral galaxies (Hodge 1969; Hodge \& Kennicutt 1983;
Athanassoula \etal 1993, Gonzalez Delgado \& Perez 1997) with normal (e.g., not
giant) HII regions.  Since the grism data and detection method used
here is optimized to find the brightest star--forming regions only, a
direct comparison to these studies is not straightforward.  However,
we examine here a subset of ten of the multiple--knot
emitters---excluding visually disturbed galaxies such as mergers and
objects with nearby companions---in order to determine the radial
distribution of HII regions.  We exclude irregular galaxies and/or
mergers---that could have emitting knots in the tidal tail, for
example---since such enhanced star formation is likely induced
predominantly by the dynamics of the interaction and not that which
normally occurs in undisturbed disk galaxies.  Such exclusion of
irregular galaxies in radial distribution studies was also done by, e.g.,
Athanassoula \etal (1993).  This subset of ten galaxies with multiple
emitting knots contains a total of 26 knots, within a redshift range
of 0.076--0.483 (seven of which are above z\cge0.1).  The galaxy knots
are distributed across the faces of the galaxies as shown in
Figure~\ref{fig:multknotsmontage}.

\begin{figure}
\includegraphics[scale=0.5]{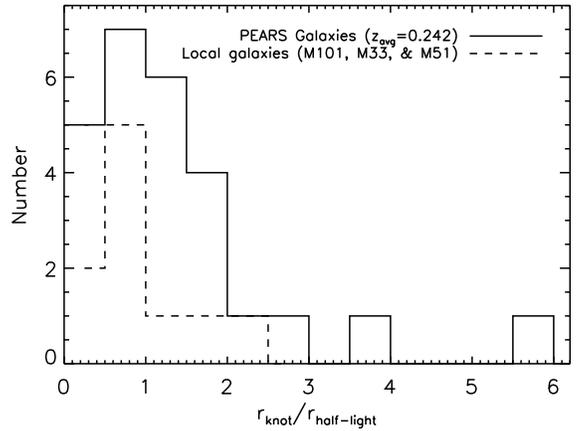}
\caption{Comparison of radial distributions of star--forming regions within the PEARS galaxies that have multiple (giant) star--forming knots to a sample of local galaxies with well--known giant HII regions.  Radial knot distances are all scaled to the half--light radii of the galaxy, as described in detail in the text.  A few of the PEARS knots shown here could be considered nuclear.  The PEARS sample of galaxies with multiple--emitting knots has an average redshift of z=0.242, and both samples peak near the half--light radius.}

\label{fig:radialdist}
\end{figure}

The distribution of PEARS galaxy knots (Figure~\ref{fig:radialdist}) peaks around the normalized half--light radii (obtained from the GOODS catalogs).  A few of these could be considered nuclear star--forming knots, as is seen in Figure~\ref{fig:multknotsmontage}.   As a comparison to the PEARS star--forming regions at an average redshift z$=$0.242, we also plot the radial distribution of well--studied giant extragalactic HII regions (Kennicutt 1984; Blitz \etal 1981; Castaneda \etal 1992; Rosa \etal 1984) in the local universe.  This sample is drawn from M101, M33, and M51.  From these relatively small samples, we see a peak in both the local and PEARS galaxies' giant star--forming regions around the half--light radius.  We note here that since the median radial distance is much larger than the ACS resolution for the PEARS galaxies, the peak is not due to resolution effects.  Since line emission from giant HII regions is a good tracer of massive star formation---which is a proxy to the galaxies' evolution---this result suggests that if there is a fundamental parameter controlling the normalized radial distance of these occurances, it persists to intermediate redshifts probed in this study.


\section{Summary}
We present results from a search for emission--line galaxies in the
five PEARS South Fields, including the HUDF.  We outline briefly the
method used to arrive at our catalog, which relies on spectral
extractions from individual emitting knots within galaxies, detected
first in the 2D grism image.  In this way, we detect emission--line
sources that would likely otherwise be missed in the standard
extraction of entire galaxies, where continuum flux can often dominate
the spectrum and wash out the line.  Here we summarize our findings:

1.  We detect 320 emission lines from 226 galaxy knots within
192 individual galaxies.  The most common emission lines are \OIII,
\Ha\, and \OII\---we detect 136, 83, and 30 emission lines of each species, respectively.
We detect 25 galaxies with multiple emitting knots.

2.  In Table 1 we present 118 new grism spectroscopic
redshifts in the GOODS--South Field.  Line identifications are
obtained by either wavelength ratios where two lines are present in a
given spectrum, or by utilizing previously measured---typically
photometric---redshifts for these objects as a first--guess.
  
3.  We calculate SFRs of the ELG sample using \Ha\ and
\OII\ where available, and derive an \OIII\ SFR based on the more
dependable lines when two lines are available in the spectra.  The SFR
as a function of redshift is given in Figure~\ref{fig:sfr}.

4.  Including blended \Hb\ in our line fits results in
identification of probable AGN based on approximate excitation levels.
In comparison to AGN from SDSS, we find that the PEARS AGN candidates
are situated in the high--excitation, high--luminosity region of the
distribution.

5.  The 2D--detection method used for the PEARS South
grism data is optimal for the detection of individual star--forming regions in
galaxies up to z$\sim$0.5.  We find that the normalized radial
distance of giant star--forming knots peaks near the half--light radii of
the galaxies---as does a comparison sample of nearby giant HII regions
in M101, M31, and M51.
  
Future work will begin with analysis of the PEARS North Fields data,
which are currently being reduced, and will result in the second in
this series of papers.  Detailed studies using sources from both the
PEARS South and North Fields will include an in--depth study of line
luminosity functions and star--formation rate densities, which will be
possible once simulations of the data are completed in order to obtain
accurate estimates of incompleteness.  Future slitless spectroscopy
studies with the Wide Field Camera 3 will provide a wealth of information and
confirmation for the objects already identified here, as well as
detection of new ELGs at higher redshifts.  Additionally, the James Webb Space Telescope (Gardner \etal 2006) will
extend the study of ELGs to still longer wavelengths---and higher
redshifts---adding vastly to our knowledge of actively
star--forming galaxies which are fundamental in the overall study of
galaxy evolution.

We thank Mark Dickinson for useful discussions.  This research was
supported in part by the NASA/UNCFSP Harriett G. Jenkins Predoctoral
Fellowship program and by an appointment to the NASA Postdoctoral
Program at Goddard Space Flight Center, administered by Oak Ridge
Associated Universities through a contract with NASA (ANS), as well as
by grants HST-GO-10530 \& HST-GO-9793 from STScI, which is operated by
AURA for NASA under contract NAS 5-26555.  We thank the anonymous referee for helpful comments which improved the paper.




\begin{longtable}{cccccccccl}
\tabletypesize{\scriptsize}
\tablecaption{Global Properties of Emission--Line Galaxies \label{table1}}
\tablewidth{0pt}
\tablehead{ \colhead{PEARS}&\colhead{Knot}&\colhead{RA}&\colhead{Dec}&\colhead{$i'_{AB}$}&\colhead{Wavelength}&\colhead{Flux}&\colhead{EW}&\colhead{Line}&\colhead{Grism \hspace{0.5cm}   Flag}\\
\colhead{ID}&\colhead{\# }&\colhead{(deg)}&\colhead{(deg)}&\colhead{(mag)}&\colhead{(\AA)}&\colhead{($10^{-18} erg/s/cm^2$)}&\colhead{(\AA)}&\colhead{ID}&\colhead{\hspace{-0.2cm}Redshift \hspace{0.5cm} (*) }
}
\startdata
9359 & 1 & 53.1682091 & -27.9300213 & 22.97 & 7349 & 21.2$\pm$3.5 & 94 & \Ha & 0.120 \hspace{0.85cm}  2 \\ 
9359 & 2 & 53.1682892 & -27.9300632 & 22.97 & 7385 & 19.7$\pm$3.5 & 75 & \Ha & 0.125   \hspace{0.85cm} 2 \\ 
12250 & 1 & 53.1566811 & -27.9257526 & 24.72 & 6671 & 57.0$\pm$2.0 & 313 & \OIII & 0.337   \hspace{0.85cm} 1 \\ 
12250 & 1 & 53.1566811 & -27.9257526 & 24.72 & 8774 & 26.1$\pm$5.3 & 199 & \Ha & 0.337   \hspace{0.85cm} 1 \\ 
12665 & 1 & 53.1541176 & -27.9234123 & 22.06 & 7413 & 137.1$\pm$6.4 & 214 & \nodata & \nodata \hspace{0.7cm}   \nodata \\ 
13541 & 1 & 53.1584473 & -27.9188538 & 21.49 & 6842 & 112.4$\pm$2.8 & 111 & \OIII & 0.370   \hspace{0.85cm} 1 \\ 
13541 & 1 & 53.1584473 & -27.9188538 & 21.49 & 9059 & 75.7$\pm$8.8 & 83 & \Ha & 0.370   \hspace{0.85cm} 1 \\ 
13541 & 2 & 53.1584740 & -27.9189358 & 21.49 & 6802 & 16.5$\pm$1.4 & 10 & \OIII & 0.370   \hspace{0.85cm} 1 \\ 
13541 & 2 & 53.1584740 & -27.9189358 & 21.49 & 9013 & 95.6$\pm$11.2 & 76 & \Ha & 0.370   \hspace{0.85cm} 1 \\ 
13553 & 1 & 53.1643639 & -27.9186115 & 19.46 & 7416 & 23.0$\pm$6.9 & 19 & \Ha & 0.130   \hspace{0.85cm} 2 \\ 
14104 & 1 & 53.1616020 & -27.9222107 & 26.60 & 8098 & 21.2$\pm$2.2 & 344 & \OIII & 0.621   \hspace{0.85cm} 3 \\ 
14215\tablenotemark{\ddag} & 1 & 53.1456490 & -27.9198093 & 22.20 & 6798 & 9.5$\pm$2.8 & 11 & \OII & 0.832   \hspace{0.85cm} 1 \\ 
14215\tablenotemark{\ddag} & 1 & 53.1456490 & -27.9198093 & 22.20 & 9153 & 37.4$\pm$6.2 & 37 & \OIII & 0.832   \hspace{0.85cm} 1 \\ 
15116 & 1 & 53.1471062 & -27.9200668 & 25.61 & 6618 & 80.4$\pm$4.4 & 291 & \OIII & 0.332   \hspace{0.85cm} 1 \\ 
15116 & 1 & 53.1471062 & -27.9200668 & 25.61 & 8742 & 22.4$\pm$3.8 & 139 & \Ha & 0.332   \hspace{0.85cm} 1 \\ 
16788 & 1 & 53.1849480 & -27.9160748 & 25.41 & 7846 & 3.6$\pm$0.9 & 28 & \Hb & 0.659   \hspace{0.85cm} 3 \\ 
16788 & 1 & 53.1849480 & -27.9160748 & 25.41 & 8065 & 17.7$\pm$1.2 & 140 & \OIII & 0.659   \hspace{0.85cm} 3 \\ 
17024 & 1 & 53.1534920 & -27.9139519 & 22.20 & 8070 & 3.8$\pm$0.9 & 13 & \Hb & 0.659   \hspace{0.85cm} 2 \\ 
17024 & 1 & 53.1534920 & -27.9139519 & 22.20 & 8288 & 18.7$\pm$1.5 & 55 & \OIII & 0.659   \hspace{0.85cm} 2 \\ 
17362 & 1 & 53.1624565 & -27.9140434 & 22.56 & 7367 & 8.1$\pm$1.7 & 13 & \OIII & 0.475   \hspace{0.85cm} 2 \\ 
17587 & 1 & 53.1607971 & -27.9138756 & 24.82 & 8240 & 21.3$\pm$1.9 & 150 & \OIII & 0.650   \hspace{0.85cm} 3 \\ 
18337 & 1 & 53.1779099 & -27.9095287 & 22.25 & 7900 & 35.0$\pm$6.0 & 24 & \OII & 1.120   \hspace{0.85cm} 2 \\ 
18410 & 1 & 53.1801987 & -27.9110069 & 22.55 & 7086 & 4.9$\pm$1.0 & 16 & \Hb & 0.454   \hspace{0.85cm} 3 \\ 
18410 & 1 & 53.1801987 & -27.9110069 & 22.55 & 7264 & 16.8$\pm$1.1 & 53 & \OIII & 0.454   \hspace{0.85cm} 3 \\ 
19422 & 1 & 53.1720695 & -27.9096584 & 24.52 & 7757 & 44.6$\pm$2.0 & 315 & \OIII & 0.553   \hspace{0.85cm} 3 \\ 
19546 & 1 & 53.1468811 & -27.9090500 & 23.61 & 9049 & 20.2$\pm$5.4 & 36 & \OII & 1.428   \hspace{0.85cm} 2 \\ 
19639 & 1 & 53.1456108 & -27.9038506 & 19.92 & 8399 & 62.0$\pm$8.6 & 57 & \Ha & 0.280   \hspace{0.85cm} 2 \\ 
20201 & 1 & 53.1496544 & -27.9078388 & 24.14 & 7217 & 186.3$\pm$5.0 & 529 & \OIII & 0.445   \hspace{0.85cm} 1 \\ 
20201 & 1 & 53.1496544 & -27.9078388 & 24.14 & 9514 & 52.7$\pm$8.0 & 231 & \Ha & 0.445   \hspace{0.85cm} 1 \\ 
21605 & 1 & 53.1766396 & -27.9037571 & 23.13 & 8156 & 35.1$\pm$4.2 & 59 & \OII & 1.188   \hspace{0.85cm} 2 \\ 
21754 & 1 & 53.1799278 & -27.9038067 & 24.07 & 7802 & 19.2$\pm$3.6 & 90 & \OIII & 0.562   \hspace{0.85cm} 3 \\ 
22203 & 1 & 53.1505966 & -27.9024887 & 22.13 & 6330 & 183.8$\pm$11.7 & 167 & \OIII & 0.267 \hspace{0.85cm}   1  \\ 
22203 & 1 & 53.1505966 & -27.9024887 & 22.13 & 8427 & 91.9$\pm$6.8 & 134 & \Ha & 0.267   \hspace{0.85cm} 1 \\ 
22604 & 1 & 53.1840134 & -27.8996315 & 20.99 & 7975 & 19.9$\pm$3.5 & 19 & \nodata & \nodata   \hspace{0.7cm} \nodata \\
22761 & 1 & 53.1536331 & -27.9019623 & 24.46 & 6394 & 18.1$\pm$4.0 & 124 & \nodata & \nodata \hspace{0.7cm} \nodata \\
22829 & 1 & 53.1647568 & -27.9002018 & 21.55 & 7787 & 74.0$\pm$3.5 & 47 & \OIII & 0.559   \hspace{0.85cm} 3 \\ 
26009 & 1 & 53.1379280 & -27.8946438 & 23.63 & 7163 & 62.0$\pm$3.5 & 203 & \OIII & 0.439   \hspace{0.85cm} 3 \\ 
26107 & 1 & 53.1397209 & -27.8947048 & 25.48 & 8443 & 36.9$\pm$3.6 & 318 & \nodata & \nodata   \hspace{0.7cm} \nodata \\ 
26909 & 1 & 53.1390152 & -27.8926620 & 23.02 & 8388 & 21.3$\pm$1.1 & 45 & \Hb & 0.679   \hspace{0.85cm} 1 \\ 
26909 & 1 & 53.1390152 & -27.8926620 & 23.02 & 8388 & 219.4$\pm$2.6 & 454 & \OIII & 0.679   \hspace{0.85cm} 1 \\ 
26009 & 1 & 53.1379280 & -27.8946438 & 23.02 & 9441 & 18.7$\pm$5.6 & 81 & \Ha & 0.439   \hspace{0.85cm} 1 \\ 
26009 & 1 & 53.1379280 & -27.8946438 & 23.63 & 7163 & 60.6$\pm$1.6 & 204 & \OIII & 0.439   \hspace{0.85cm} 1 \\ 
27293 & 1 & 53.1968880 & -27.8927078 & 25.89 & 7693 & 18.1$\pm$4.6 & 304 & \nodata & \nodata   \hspace{0.7cm} \nodata \\ 
30908 & 1 & 53.1774559 & -27.8812027 & 20.39 & 7234 & 38.2$\pm$5.1 & 15 & \nodata & \nodata   \hspace{0.7cm} \nodata \\ 
31362 & 1 & 53.1819839 & -27.8849945 & 24.22 & 8096 & 15.8$\pm$3.3 & 50 & \Hb & 0.665   \hspace{0.85cm} 3 \\ 
31362 & 1 & 53.1819839 & -27.8849945 & 24.22 & 8318 & 171.8$\pm$3.7 & 612 & \OIII & 0.665   \hspace{0.85cm} 3 \\ 
32905 & 1 & 53.1583290 & -27.8816776 & 23.89 & 7451 & 12.9$\pm$2.4 & 58 & \OII & 0.999   \hspace{0.85cm} 3 \\ 
33086\tablenotemark{\ddag} & 1 & 53.1846313 & -27.8809357 & 23.89 & 6891 & 80.2$\pm$4.5 & 53 & CIII] & 3.446   \hspace{0.85cm} 1 \\ 
33086\tablenotemark{\ddag} & 1 & 53.1846313 & -27.8809357 & 23.89 & 8490 & 21.5$\pm$4.6 & 25 & CIV & 3.446   \hspace{0.85cm} 1 \\ 
33294 & 1 & 53.1586571 & -27.8801899 & 23.51 & 7630 & 14.1$\pm$3.1 & 29 & \OII & 1.047   \hspace{0.85cm} 2 \\ 
33355 & 1 & 53.1859474 & -27.8796978 & 23.19 & 7083 & 47.7$\pm$3.6 & 176 & \nodata & \nodata   \hspace{0.7cm} \nodata \\ 
35818 & 1 & 53.1835709 & -27.8625374 & 18.86 & 8392 & 115.6$\pm$13.8 & 71 & \Ha & 0.279 \hspace{0.85cm}   4  \\ 
36657\tablenotemark{\ddag} & 1 & 53.1743889 & -27.8673668 & 22.41 & 7141 & 180.6$\pm$12.6 & 30 & CIII] & 3.607   \hspace{0.85cm} 1 \\ 
36657\tablenotemark{\ddag} & 1 & 53.1743889 & -27.8673668 & 22.41 & 8759 & 110.8$\pm$11.1 & 25 & CIV & 3.607   \hspace{0.85cm} 1 \\ 
37690 & 1 & 53.1697388 & -27.8713989 & 23.61 & 6793 & 16.1$\pm$4.7 & 50 & \nodata & \nodata   \hspace{0.7cm} \nodata \\ 
38750 & 1 & 53.1655579 & -27.8651085 & 20.82 & 6871 & 17.2$\pm$1.6 & 17 & \Hb & 0.408   \hspace{0.85cm} 1 \\ 
38750 & 1 & 53.1655579 & -27.8651085 & 20.82 & 7032 & 85.2$\pm$3.0 & 69 & \OIII & 0.408   \hspace{0.85cm} 1 \\ 
38750 & 1 & 53.1655579 & -27.8651085 & 20.82 & 9345 & 93.9$\pm$12.9 & 87 & \Ha & 0.408   \hspace{0.85cm} 1 \\ 
39387 & 1 & 53.1705093 & -27.8666763 & 21.31 & 6725 & 26.3$\pm$4.8 & 63 & \OIII & 0.353   \hspace{0.85cm} 1 \\ 
39387 & 1 & 53.1705093 & -27.8666763 & 21.31 & 8881 & 37.3$\pm$10.7 & 30 & \Ha & 0.353   \hspace{0.85cm} 1 \\ 
40117 & 1 & 53.1818123 & -27.8668137 & 23.96 & 8520 & 25.6$\pm$4.3 & 143 & \OIII & 0.706   \hspace{0.85cm} 3 \\ 
40163 & 1 & 53.1791611 & -27.8665504 & 24.06 & 9474 & 19.2$\pm$1.4 & 61 & \OIII & 0.897   \hspace{0.85cm} 2 \\ 
40816\tablenotemark{\dag} & 1 & 53.1844559 & -27.8614178 & 19.48 & 6182 & 56.0$\pm$17.0 & 13 & \Hb & 0.281   \hspace{0.85cm} 1 \\ 
40816\tablenotemark{\dag} & 1 & 53.1844559 & -27.8614178 & 19.48 & 6353 & 706.6$\pm$21.7 & 147 & \OIII & 0.281   \hspace{0.85cm} 1 \\ 
40816\tablenotemark{\dag} & 1 & 53.1844559 & -27.8614178 & 19.48 & 8408 & 967.9$\pm$41.2 & 161 & \Ha & 0.281   \hspace{0.85cm} 1 \\ 
41078 & 1 & 53.1807861 & -27.8651638 & 24.30 & 6953 & 8.9$\pm$2.9 & 29 & \OII & 0.866   \hspace{0.85cm} 1 \\ 
41078 & 1 & 53.1807861 & -27.8651638 & 24.30 & 9316 & 38.0$\pm$7.2 & 130 & \OIII & 0.866   \hspace{0.85cm} 1 \\ 
43170 & 1 & 53.1561699 & -27.8608074 & 24.06 & 8451 & 53.6$\pm$4.6 & 348 & \OIII & 0.692   \hspace{0.85cm} 3 \\ 
45223 & 1 & 53.1973305 & -27.8572559 & 24.48 & 8106 & 5.3$\pm$1.2 & 27 & \Hb & 0.668   \hspace{0.85cm} 3 \\ 
45223 & 1 & 53.1973305 & -27.8572559 & 24.48 & 8331 & 42.4$\pm$2.0 & 207 & \OIII & 0.668   \hspace{0.85cm} 3 \\ 
45454 & 1 & 53.1814690 & -27.8563213 & 22.75 & 7105 & 9.7$\pm$0.9 & 81 & \OIII & 0.425   \hspace{0.85cm} 1 \\ 
46562 & 1 & 53.1642036 & -27.8534393 & 23.03 & 8319 & 22.7$\pm$2.0 & 82 & \OIII & 0.665   \hspace{0.85cm} 3 \\* 
46994 & 1 & 53.1643944 & -27.8536777 & 24.22 & 8108 & 7.2$\pm$1.0 & 32 & \Hb & 0.668   \hspace{0.85cm} 3 \\ 
46994 & 1 & 53.1643944 & -27.8536777 & 24.22 & 8332 & 72.7$\pm$1.9 & 287 & \OIII & 0.668   \hspace{0.85cm} 3 \\ 
48890 & 1 & 53.2069473 & -27.8480740 & 23.07 & 7110 & 15.7$\pm$3.5 & 19 & \OII & 0.908   \hspace{0.85cm} 1 \\ 
48890 & 1 & 53.2069473 & -27.8480740 & 23.07 & 9566 & 38.9$\pm$12.2 & 49 & \OIII & 0.908   \hspace{0.85cm} 1 \\ 
49766 & 1 & 53.1751556 & -27.8476467 & 23.57 & 6059 & 20.3$\pm$5.1 & 285 & \OIII & 0.213   \hspace{0.85cm} 1 \\ 
49766 & 1 & 53.1751556 & -27.8476467 & 23.57 & 8041 & 11.6$\pm$3.3 & 65 & \Ha & 0.213   \hspace{0.85cm} 1 \\ 
51061 & 1 & 53.2095871 & -27.8468800 & 25.50 & 7890 & 20.5$\pm$3.2 & 173 & \OIII & 0.580   \hspace{0.85cm} 3 \\ 
51356 & 1 & 53.1567993 & -27.8429546 & 21.59 & 8126 & 20.5$\pm$2.1 & 79 & \OIII & 0.627   \hspace{0.85cm} 4 \\ 
51976 & 1 & 53.2045479 & -27.8425064 & 23.73 & 9280 & 56.0$\pm$10.2 & 147 & \OIII & 0.862   \hspace{0.85cm} 1 \\ 
51976 & 1 & 53.2045479 & -27.8425064 & 23.73 & 6939 & 14.8$\pm$4.0 & 43 & \OII & 0.86   \hspace{0.85cm} 1 \\ 
52086 & 1 & 53.1578102 & -27.8444538 & 23.53 & 4954 & 4.1$\pm$0.7 & 11 & \Hb & 0.526   \hspace{0.85cm} 3 \\ 
52086 & 1 & 53.1578102 & -27.8444538 & 23.53 & 7620 & 12.4$\pm$1.0 & 127 & \OIII & 0.526   \hspace{0.85cm} 3 \\ 
52086 & 2 & 53.1577988 & -27.8443146 & 23.53 & 7419 & 6.6$\pm$1.0 & 19 & \Hb & 0.526   \hspace{0.85cm} 3 \\ 
52086 & 2 & 53.1577988 & -27.8443146 & 23.53 & 7619 & 117.3$\pm$3.2 & 317 & \OIII & 0.526   \hspace{0.85cm} 3 \\ 
52398 & 1 & 53.1938133 & -27.8442917 & 24.19 & 8323 & 12.7$\pm$3.3 & 38 & \nodata & \nodata   \hspace{0.7cm} \nodata \\ 
52502 & 1 & 53.2011223 & -27.8412380 & 21.53 & 7216 & 28.1$\pm$1.7 & 84 & \OIII & 0.445   \hspace{0.85cm} 1 \\ 
52502 & 1 & 53.2011223 & -27.8412380 & 21.53 & 9485 & 29.9$\pm$9.0 & 56 & \Ha & 0.445   \hspace{0.85cm} 1 \\ 
53462 & 1 & 53.1896210 & -27.8419609 & 22.81 & 7804 & 18.9$\pm$3.8 & 40 & \OIII & 0.562   \hspace{0.85cm} 2 \\ 
54022 & 1 & 53.1747398 & -27.8407822 & 22.37 & 6669 & 18.0$\pm$2.3 & 41 & \OIII & 0.336   \hspace{0.85cm} 1 \\ 
54022 & 1 & 53.1747398 & -27.8407822 & 22.37 & 8769 & 20.0$\pm$5.6 & 62 & \Ha & 0.336   \hspace{0.85cm} 1 \\ 
54022 & 2 & 53.1746254 & -27.8408928 & 22.37 & 6499 & 4.2$\pm$1.4 & 15 & \Hb & 0.336   \hspace{0.85cm} 1 \\ 
54022 & 2 & 53.1746254 & -27.8408928 & 22.37 & 6675 & 34.7$\pm$3.4 & 94 & \OIII & 0.336   \hspace{0.85cm} 1 \\ 
55102 & 1 & 53.1756477 & -27.8385963 & 21.85 & 7224 & 30.2$\pm$5.3 & 62 & \OIII & 0.458   \hspace{0.85cm} 2 \\ 
55102 & 2 & 53.1757545 & -27.8387928 & 21.85 & 7281 & 34.5$\pm$1.6 & 134 & \OIII & 0.458   \hspace{0.85cm} 2 \\ 
56801 & 1 & 53.1450958 & -27.8373871 & 23.96 & 8040 & 7.5$\pm$0.9 & 33 & \Hb & 0.653   \hspace{0.85cm} 3 \\ 
56801 & 1 & 53.1450958 & -27.8373871 & 23.96 & 8259 & 80.2$\pm$3.4 & 297 & \OIII & 0.653   \hspace{0.85cm} 3 \\ 
56875 & 1 & 53.1529732 & -27.8376999 & 24.52 & 7695 & 11.7$\pm$1.2 & 99 & \OIII & 0.541   \hspace{0.85cm} 3 \\ 
58985 & 1 & 53.1999168 & -27.8340626 & 23.78 & 7807 & 23.2$\pm$3.4 & 102 & \OIII & 0.563   \hspace{0.85cm} 3 \\ 
59018 & 1 & 53.1763496 & -27.8306465 & 20.63 & 9606 & 56.5$\pm$13.2 & 40 & \Ha & 0.464   \hspace{0.85cm} 2 \\ 
59905 & 1 & 53.1802177 & -27.8329601 & 25.30 & 8417 & 35.8$\pm$4.2 & 186 & \nodata & \nodata   \hspace{0.7cm} \nodata \\ 
60143 & 1 & 53.1482964 & -27.8289299 & 21.25 & 7729 & 30.8$\pm$3.5 & 37 & \OIII & 0.542   \hspace{0.85cm} 2 \\ 
63612 & 1 & 53.1554031 & -27.8261337 & 25.33 & 7820 & 3.3$\pm$0.7 & 27 & \Hb & 0.606   \hspace{0.85cm} 3 \\ 
63612 & 1 & 53.1554031 & -27.8261337 & 25.33 & 8022 & 12.4$\pm$1.1 & 128 & \OIII & 0.606   \hspace{0.85cm} 3 \\ 
66061\tablenotemark{\ddag} & 1 & 53.1801453 & -27.8206253 & 22.68 & 8161 & 81.5$\pm$6.7 & 74 & MgII & 1.917   \hspace{0.85cm} 2 \\ 
68739 & 1 & 53.1607819 & -27.8163223 & 24.88 & 7624 & 11.9$\pm$1.0 & 162 & \OIII & 0.526   \hspace{0.85cm} 3 \\ 
70314 & 1 & 53.1748161 & -27.7995949 & 20.36 & 7526 & 28.4$\pm$13.9 & 69 & \Ha & 0.147   \hspace{0.85cm} 2 \\ 
70314 & 2 & 53.1750183 & -27.7993336 & 20.36 & 7527 & 29.1$\pm$5.6 & 86 & \Ha & 0.147   \hspace{0.85cm} 2 \\ 
70337 & 1 & 53.1938210 & -27.8128853 & 23.40 & 6802 & 20.3$\pm$3.5 & 60 & \OII & 0.830   \hspace{0.85cm} 1 \\ 
70337 & 1 & 53.1938210 & -27.8128853 & 23.40 & 9142 & 38.7$\pm$6.7 & 106 & \OIII & 0.830   \hspace{0.85cm} 1 \\ 
70407\tablenotemark{\dag} & 1 & 53.1851540 & -27.8052826 & 20.52 & 9300 & 60.4$\pm$26.1 & 28 & \Ha & 0.417   \hspace{0.85cm} 2 \\ 
70407\tablenotemark{\dag} & 2 & 53.1851730 & -27.8052406 & 20.52 & 9452 & 42.4$\pm$20.0 & 20 & \Ha & 0.417   \hspace{0.85cm} 2 \\ 
70651 & 1 & 53.1530495 & -27.8121529 & 23.33 & 6040 & 80.0$\pm$9.8 & 559 & \OIII & 0.212   \hspace{0.85cm} 1 \\ 
70651 & 1 & 53.1530495 & -27.8121529 & 23.33 & 7956 & 30.4$\pm$3.3 & 296 & \Ha & 0.212   \hspace{0.85cm} 1 \\ 
71864 & 1 & 53.1506119 & -27.8095131 & 24.86 & 8786 & 13.2$\pm$0.7 & 80 & \OIII & 0.759   \hspace{0.85cm} 3 \\ 
71924 & 1 & 53.1536827 & -27.8088989 & 23.84 & 6898 & 3.6$\pm$1.0 & 13 & \OIII & 0.381   \hspace{0.85cm} 3 \\ 
72179 & 1 & 53.1310501 & -27.8084450 & 23.34 & 6281 & 29.1$\pm$10.5 & 94 & \OIII & 0.257   \hspace{0.85cm} 3 \\ 
72509 & 1 & 53.1705208 & -27.8066082 & 24.53 & 8548 & 7.1$\pm$3.1 & 28 & \OII & 1.294   \hspace{0.85cm} 2 \\ 
72557 & 1 & 53.1338768 & -27.8068733 & 23.56 & 6673 & 15.7$\pm$3.2 & 477 & \nodata & \nodata   \hspace{0.7cm} \nodata \\ 
73619 & 1 & 53.1844063 & -27.8051853 & 26.88 & 8250 & 10.1$\pm$1.2 & 69 & \OIII & 0.652   \hspace{0.85cm} 3 \\ 
74234 & 1 & 53.1377335 & -27.8042202 & 25.95 & 7492 & 2.7$\pm$0.7 & 35 & \Hb & 0.542   \hspace{0.85cm} 3 \\ 
74234 & 1 & 53.1377335 & -27.8042202 & 25.95 & 7705 & 16.8$\pm$1.4 & 207 & \OIII & 0.542   \hspace{0.85cm} 3 \\ 
75506 & 1 & 53.1472664 & -27.8008537 & 26.26 & 8419 & 17.1$\pm$7.3 & 1158 & \Ha & 0.277   \hspace{0.85cm} 1 \\ 
75506 & 1 & 53.1472664 & -27.8008537 & 26.26 & 6379 & 12.1$\pm$0.9 & 162 & \OIII & 0.277   \hspace{0.85cm} 1 \\ 
75547 & 1 & 53.1733017 & -27.7993031 & 23.78 & 7372 & 7.8$\pm$2.4 & 60 & \Ha & 0.123   \hspace{0.85cm} 3 \\ 
75547 & 2 & 53.1732941 & -27.7992783 & 23.78 & 7372 & 6.3$\pm$1.9 & 46 & \Ha & 0.123   \hspace{0.85cm} 3 \\ 
75753 & 1 & 53.1872597 & -27.7943401 & 22.29 & 6499 & 7.2$\pm$1.1 & 20 & \Hb & 0.343   \hspace{0.85cm} 1 \\ 
75753 & 1 & 53.1872597 & -27.7943401 & 22.29 & 6709 & 61.3$\pm$1.5 & 143 & \OIII & 0.343   \hspace{0.85cm} 1 \\ 
75753 & 1 & 53.1872597 & -27.7943401 & 22.29 & 8816 & 32.8$\pm$4.2 & 180 & \Ha & 0.343   \hspace{0.85cm} 1 \\ 
75753 & 2 & 53.1873703 & -27.7942238 & 22.29 & 6490 & 17.3$\pm$0.5 & 29 & \Hb & 0.343   \hspace{0.85cm} 1 \\ 
75753 & 2 & 53.1873703 & -27.7942238 & 22.29 & 6685 & 152.8$\pm$1.0 & 228 & \OIII & 0.343   \hspace{0.85cm} 1 \\ 
75753 & 2 & 53.1873703 & -27.7942238 & 22.29 & 8819 & 65.5$\pm$4.9 & 208 & \Ha & 0.343   \hspace{0.85cm} 1 \\ 
75753 & 3 & 53.1878090 & -27.7939053 & 22.29 & 8800 & 80.0$\pm$6.9 & 124 & \Ha & 0.343   \hspace{0.85cm} 1 \\ 
75753 & 3 & 53.1878090 & -27.7939053 & 22.29 & 6697 & 186.0$\pm$3.9 & 234 & \OIII & 0.343   \hspace{0.85cm} 1 \\ 
76154 & 1 & 53.1512299 & -27.7987995 & 23.73 & 8016 & 39.9$\pm$1.3 & 183 & \OIII & 0.600   \hspace{0.85cm} 3 \\ 
77558 & 1 & 53.1864052 & -27.7910328 & 18.67 & 7995 & 23.1$\pm$5.2 & \nodata & \nodata & \nodata   \hspace{0.7cm} \nodata \\ 
77558 & 2 & 53.1871910 & -27.7909679 & 18.67 & 7894 & 84.6$\pm$15.1 & \nodata & \nodata & \nodata   \hspace{0.7cm} \nodata \\ 
77902 & 1 & 53.1559830 & -27.7949619 & 23.51 & 7720 & 7.4$\pm$2.3 & 53 & \OII & 1.071   \hspace{0.85cm} 3 \\ 
78021 & 1 & 53.1839218 & -27.7954350 & 27.62 & 8615 & 13.3$\pm$3.6 & 294 & \OII & 1.311   \hspace{0.85cm} 3 \\ 
78077 & 1 & 53.1841545 & -27.7926388 & 21.73 & 6482 & 48.2$\pm$17.7 & 32 & \OII & 0.737   \hspace{0.85cm} 1 \\ 
78077 & 1 & 53.1841545 & -27.7926388 & 21.73 & 8675 & 69.7$\pm$4.8 & 42 & \OIII & 0.737   \hspace{0.85cm} 1 \\ 
78237 & 1 & 53.1876869 & -27.7943954 & 20.50 & 6693 & 14.4$\pm$0.9 & 26 & \OIII & 0.340   \hspace{0.85cm} 1 \\ 
78237 & 1 & 53.1876869 & -27.7943954 & 20.50 & 8800 & 38.7$\pm$6.8 & 160 & \Ha & 0.340   \hspace{0.85cm} 1 \\ 
78237 & 2 & 53.1879768 & -27.7943249 & 20.50 & 6701 & 70.2$\pm$16.6 & 382 & \OIII & 0.340   \hspace{0.85cm} 1 \\ 
78237 & 3 & 53.1877136 & -27.7942200 & 20.50 & 7872 & 20.6$\pm$3.8 & 40 & \nodata & \nodata   \hspace{0.7cm} \nodata \\ 
78491 & 1 & 53.1548195 & -27.7934532 & 22.70 & 7143 & 15.4$\pm$3.9 & 237 & \nodata & \nodata   \hspace{0.7cm} \nodata \\ 
78491 & 2 & 53.1544189 & -27.7933788 & 22.70 & 8100 & 10.3$\pm$2.7 & 224 & \Ha & 0.234   \hspace{0.85cm} 1 \\ 
78491 & 2 & 53.1544189 & -27.7933788 & 22.70 & 6120 & 17.2$\pm$3.8 & 190 & \OIII & 0.234   \hspace{0.85cm} 1 \\ 
78491 & 3 & 53.1547127 & -27.7931709 & 22.70 & 6114 & 54.5$\pm$5.4 & 276 & \OIII & 0.234   \hspace{0.85cm} 1 \\ 
78491 & 3 & 53.1547127 & -27.7931709 & 22.70 & 8080 & 20.4$\pm$3.0 & 144 & \Ha & 0.234   \hspace{0.85cm} 1 \\ 
78582\tablenotemark{\dag} & 1 & 53.1615829 & -27.7922630 & 21.18 & 7013 & 12.3$\pm$2.4 & 6 & \Hb & 0.454   \hspace{0.85cm} 1 \\ 
78582\tablenotemark{\dag} & 1 & 53.1615829 & -27.7922630 & 21.18 & 7100 & 39.5$\pm$2.4 & 17 & \OIII & 0.454   \hspace{0.85cm} 1 \\ 
78582\tablenotemark{\dag} & 1 & 53.1615829 & -27.7922630 & 21.18 & 9542 & 276.2$\pm$54.4 & 189 & \Ha & 0.454   \hspace{0.85cm} 1 \\ 
78762 & 1 & 53.1618958 & -27.7925568 & 22.77 & 7284 & 19.2$\pm$1.0 & 83 & \OIII & 0.458   \hspace{0.85cm} 3 \\ 
79283 & 1 & 53.1419983 & -27.7867641 & 20.76 & 8070 & 19.3$\pm$5.9 & 36 & \Ha & 0.230   \hspace{0.85cm} 3 \\ 
79283 & 2 & 53.1421967 & -27.7865429 & 20.76 & 8059 & 37.7$\pm$5.9 & \nodata & \Ha & 0.230   \hspace{0.85cm} 3 \\ 
79400 & 1 & 53.1673317 & -27.7918015 & 23.98 & 6867 & 4.2$\pm$0.9 & 16 & \OIII & 0.375   \hspace{0.85cm} 3 \\ 
79483\tablenotemark{\dag} & 1 & 53.1879654 & -27.7900734 & 23.98 & 9200 & 94.9$\pm$17.6 & 54 & \Ha & 0.438   \hspace{0.85cm} 1 \\ 
79483\tablenotemark{\dag} & 2 & 53.1879539 & -27.7900009 & 23.98 & 9437 & 130.6$\pm$23.6 & 56 & \Ha & 0.438   \hspace{0.85cm} 1 \\ 
79483\tablenotemark{\dag} & 2 & 53.1879539 & -27.7900009 & 20.85 & 7001 & 5.0$\pm$1.4 & 3 & \OIII & 0.438   \hspace{0.85cm} 1 \\ 
79520 & 1 & 53.1861954 & -27.7916622 & 26.73 & 8703 & 11.3$\pm$1.4 & 56 & \OIII & 0.742   \hspace{0.85cm} 3 \\ 
80071 & 1 & 53.1866226 & -27.7902203 & 27.06 & 7335 & 47.2$\pm$3.0 & 224 & \Ha & 0.118   \hspace{0.85cm} 3 \\ 
80255 & 1 & 53.1848145 & -27.7899342 & 23.68 & 7278 & 5.5$\pm$1.6 & \nodata & \OII & 0.953   \hspace{0.85cm} 3 \\ 
80500 & 1 & 53.1472092 & -27.7884693 & 23.36 & 8064 & 4.0$\pm$0.8 & 15 & \Hb & 0.658   \hspace{0.85cm} 1 \\ 
80500 & 1 & 53.1472092 & -27.7884693 & 23.36 & 8300 & 21.2$\pm$1.1 & 83 & \OIII & 0.658   \hspace{0.85cm} 1 \\ 
80500 & 1 & 53.1472092 & -27.7884693 & 23.36 & 6178 & 11.8$\pm$3.5 & 43 & \OII & 0.658   \hspace{0.85cm} 1 \\ 
80666 & 1 & 53.1765137 & -27.7897243 & 25.04 & 6849 & 3.0$\pm$0.7 & 25 & \Hb & 0.411   \hspace{0.85cm} 3 \\ 
80666 & 1 & 53.1765137 & -27.7897243 & 25.04 & 7047 & 24.0$\pm$1.3 & 223 & \OIII & 0.411   \hspace{0.85cm} 3 \\ 
81032 & 1 & 53.1815071 & -27.7879314 & 23.36 & 6045 & 34.1$\pm$9.8 & 119 & \OIII & 0.210   \hspace{0.85cm} 2 \\ 
81256 & 1 & 53.1920815 & -27.7872849 & 23.05 & 7841 & 7.9$\pm$2.4 & 36 & \OII & 1.104   \hspace{0.85cm} 3 \\ 
81609 & 1 & 53.1640930 & -27.7872963 & 24.37 & 7820 & 15.7$\pm$2.8 & 61 & \OII & 1.098   \hspace{0.85cm} 3 \\ 
81944 & 1 & 53.1446838 & -27.7855377 & 22.53 & 8138 & 62.1$\pm$4.5 & 75 & \Ha & 0.228   \hspace{0.85cm} 1 \\ 
81944 & 2 & 53.1447372 & -27.7854137 & 22.53 & 6132 & 401.0$\pm$12.5 & 483 & \OIII & 0.228   \hspace{0.85cm} 1 \\ 
81944 & 2 & 53.1447372 & -27.7854137 & 22.53 & 8129 & 162.1$\pm$9.4 & 341 & \Ha & 0.228   \hspace{0.85cm} 1 \\ 
82307 & 1 & 53.1634598 & -27.7866497 & 25.28 & 7369 & 15.5$\pm$2.9 & \nodata & \OIII & 0.475   \hspace{0.85cm} 3 \\ 
83381 & 1 & 53.1765251 & -27.7825947 & 24.96 & 6640 & 25.3$\pm$1.5 & 293 & \OIII & 0.329   \hspace{0.85cm} 3 \\ 
83553\tablenotemark{\ddag} & 1 & 53.1784821 & -27.7840424 & 24.89 & 6462 & 94.9$\pm$5.2 & 93 & CIV & 3.166   \hspace{0.85cm} 1 \\ 
83553\tablenotemark{\ddag} & 1 & 53.1784821 & -27.7840424 & 24.89 & 7940 & 13.3$\pm$3.3 & 27 & CIII & 3.166   \hspace{0.85cm} 1 \\ 
83686 & 1 & 53.1518135 & -27.7829018 & 23.50 & 6848 & 8.9$\pm$3.1 & 39 & \OII & 0.837   \hspace{0.85cm} 3 \\ 
83789 & 1 & 53.1527901 & -27.7826843 & 24.81 & 8862 & 36.8$\pm$1.3 & 157 & \OIII & 0.774   \hspace{0.85cm} 2 \\ 
83804 & 1 & 53.1845818 & -27.7833576 & 25.04 & 7918 & 9.0$\pm$2.6 & 70 & \OII & 1.125   \hspace{0.85cm} 3 \\ 
83834 & 1 & 53.1580925 & -27.7812119 & 21.95 & 8102 & 10.8$\pm$1.1 & 34 & \OIII & 0.622   \hspace{0.85cm} 3 \\ 
85517 & 1 & 53.1763344 & -27.7808685 & 24.85 & 7409 & 3.1$\pm$0.7 & 21 & \Hb & 0.530   \hspace{0.85cm} 1 \\ 
85517 & 1 & 53.1763344 & -27.7808685 & 24.85 & 7645 & 25.0$\pm$0.9 & 175 & \OIII & 0.530   \hspace{0.85cm} 1 \\ 
85844 & 1 & 53.1624680 & -27.7803612 & 26.19 & 8569 & 22.4$\pm$3.8 & 241 & \OII & 1.299   \hspace{0.85cm} 3 \\ 
87294 & 1 & 53.1629181 & -27.7752514 & 21.03 & 8192 & 8.0$\pm$2.1 & 194 & \Ha & 0.248   \hspace{0.85cm} 3 \\ 
87464 & 1 & 53.1878166 & -27.7726479 & 22.58 & 7460 & 61.7$\pm$9.2 & 171 & \Ha & 0.130   \hspace{0.85cm} 1 \\ 
87464 & 1 & 53.1878166 & -27.7726479 & 22.58 & 5642 & 123.1$\pm$20.7 & 395 & \OIII & 0.130   \hspace{0.85cm} 1 \\ 
87658 & 1 & 53.1477661 & -27.7769241 & 24.06 & 7854 & 7.8$\pm$2.4 & 31 & \OII & 1.107   \hspace{0.85cm} 2 \\ 
88580 & 1 & 53.1620064 & -27.7740345 & 22.65 & 6354 & 10.3$\pm$0.5 & 29 & \OIII & 0.269   \hspace{0.85cm} 3 \\ 
88580 & 2 & 53.1619110 & -27.7738514 & 22.65 & 6338 & 30.5$\pm$1.1 & 87 & \OIII & 0.269   \hspace{0.85cm} 3 \\ 
89030 & 1 & 53.1604347 & -27.7752380 & 21.74 & 9127 & 32.3$\pm$4.3 & 101 & \OII & 1.449   \hspace{0.85cm} 3 \\ 
89209 & 1 & 53.1503944 & -27.7720318 & 21.27 & 6075 & 26.2$\pm$4.7 & 182 & \OIII & 0.216   \hspace{0.85cm} 1 \\ 
89209 & 1 & 53.1503944 & -27.7720318 & 21.27 & 8000 & 26.2$\pm$4.7 & 182 & \Ha & 0.216   \hspace{0.85cm} 1 \\ 
89853 & 1 & 53.1375847 & -27.7691345 & 21.91 & 8952 & 17.3$\pm$5.4 & 20 & \Ha & 0.364   \hspace{0.85cm} 2 \\ 
89923 & 1 & 53.1739769 & -27.7720718 & 21.27 & 8751 & 37.8$\pm$6.9 & 44 & \Ha & 0.333   \hspace{0.85cm} 2 \\ 
90116 & 1 & 53.1948090 & -27.7733440 & 25.48 & 8143 & 23.0$\pm$3.5 & 325 & \OIII & 0.630   \hspace{0.85cm} 3 \\ 
90246 & 1 & 53.1512070 & -27.7728481 & 24.08 & 8008 & 20.0$\pm$0.9 & 128 & \OIII & 0.603   \hspace{0.85cm} 3 \\ 
91205 & 1 & 53.1505280 & -27.7713089 & 21.27 & 7857 & 26.6$\pm$7.5 & 73 & \nodata & \nodata   \hspace{0.7cm} \nodata \\ 
91789 & 1 & 53.1470642 & -27.7701302 & 23.84 & 7655 & 8.7$\pm$0.8 & 56 & \OIII & 0.533   \hspace{0.85cm} 3 \\ 
92839\tablenotemark{\ddag} & 1 & 53.1628647 & -27.7671719 & 20.98 & 6201 & \nodata & \nodata & MgII & 1.215   \hspace{0.85cm} 2 \\ 
94632 & 1 & 53.1795502 & -27.7662010 & 25.05 & 8310 & 14.1$\pm$1.0 & 122 & \OIII & 0.664   \hspace{0.85cm} 3 \\ 
95471 & 1 & 53.1773453 & -27.7639313 & 22.40 & 8002 & 23.2$\pm$8.4 & 51 & \Ha & 0.219   \hspace{0.85cm} 2 \\ 
96123 & 1 & 53.1429062 & -27.7636814 & 23.15 & 7665 & 9.1$\pm$2.7 & 22 & \OIII & 0.535   \hspace{0.85cm} 2 \\ 
96627 & 1 & 53.1704559 & -27.7614193 & 21.53 & 7453 & 30.7$\pm$8.7 & 52 & \Ha & 0.136   \hspace{0.85cm} 2 \\ 
97568 & 1 & 53.1252556 & -27.7565651 & 22.31 & 9807 & 40.6$\pm$17.3 & 35 & \OIII & 0.963   \hspace{0.85cm} 2 \\ 
97655 & 1 & 53.1140289 & -27.7612534 & 23.72 & 7502 & 9.9$\pm$2.5 & 31 & \Hb & 0.543   \hspace{0.85cm} 3 \\ 
97655 & 1 & 53.1140289 & -27.7612534 & 23.72 & 7706 & 100.3$\pm$2.2 & 347 & \OIII & 0.543   \hspace{0.85cm} 3 \\ 
100188 & 1 & 53.1012573 & -27.7568226 & 25.07 & 6353 & 7.2$\pm$1.3 & 47 & \Hb & 0.311   \hspace{0.85cm} 1 \\ 
100188 & 1 & 53.1012573 & -27.7568226 & 25.07 & 6521 & 48.4$\pm$1.9 & 326 & \OIII & 0.311   \hspace{0.85cm} 1 \\ 
100188 & 1 & 53.1012573 & -27.7568226 & 25.07 & 8603 & 32.3$\pm$3.9 & 426 & \Ha & 0.311   \hspace{0.85cm} 1 \\ 
102156\tablenotemark{\ddag} & 1 & 53.1258965 & -27.7512856 & 21.72 & 7551 & 31.2$\pm$7.1 & 14 & NeIII & 0.738   \hspace{0.85cm} 1 \\ 
102156\tablenotemark{\ddag} & 1 & 53.1258965 & -27.7512856 & 21.72 & 8553 & 182.0$\pm$15.5 & 67 & H$\gamma$ & 0.738   \hspace{0.85cm} 1 \\ 
103116 & 1 & 53.1055984 & -27.7507782 & 22.76 & 7369 & 23.0$\pm$3.9 & 32 & \OII & 0.977   \hspace{0.85cm} 1 \\ 
103116 & 1 & 53.1055984 & -27.7507782 & 22.76 & 9900 & 59.3$\pm$4.9 & 66 & \OIII & 0.977   \hspace{0.85cm} 1 \\ 
103422 & 1 & 53.1073837 & -27.7498055 & 23.01 & 6826 & 20.4$\pm$3.9 & 28 & \OII & 0.832   \hspace{0.85cm} 1 \\ 
103422 & 1 & 53.1073837 & -27.7498055 & 23.01 & 9141 & 25.1$\pm$1.2 & 36 & \OIII & 0.832   \hspace{0.85cm} 1 \\ 
104408 & 1 & 53.1160088 & -27.7471771 & 24.26 & 8677 & 25.4$\pm$4.8 & 146 & \OIII & 0.737   \hspace{0.85cm} 3 \\ 
104849\tablenotemark{\dag} & 1 & 53.1243820 & -27.7399654 & 16.60 & 7042 & 213.3$\pm$39.5 & 27 & \Ha & 0.076   \hspace{0.85cm} 2 \\ 
104849\tablenotemark{\dag} & 2 & 53.1246300 & -27.7395153 & 16.60 & 7060 & 57.1$\pm$12.2 & 18 & \Ha & 0.076   \hspace{0.85cm} 2 \\ 
104849\tablenotemark{\dag} & 3 & 53.1243668 & -27.7394562 & 16.60 & 7069 & 101.8$\pm$19.7 & 22 & \Ha & 0.076   \hspace{0.85cm} 2 \\ 
104850\tablenotemark{\dag} & 1 & 53.1249161 & -27.7347584 & 17.02 & 7040 & 333.0$\pm$105.2 & 18 & \Ha & 0.078   \hspace{0.85cm} 2 \\ 
104850\tablenotemark{\dag} & 2 & 53.1253090 & -27.7345448 & 17.02 & 7092 & 91.1$\pm$20.5 & 18 & \Ha & 0.078   \hspace{0.85cm} 2 \\ 
104850\tablenotemark{\dag} & 3 & 53.1246567 & -27.7350121 & 17.02 & 7035 & 162.2$\pm$27.7 & 34 & \Ha & 0.078   \hspace{0.85cm} 2 \\ 
104992 & 1 & 53.1182327 & -27.7405071 & 19.65 & 7078 & 36.2$\pm$7.8 & 37 & \Ha & 0.077   \hspace{0.85cm} 2 \\ 
104992 & 2 & 53.1179123 & -27.7407436 & 19.65 & 7081 & 24.6$\pm$4.7 & 33 & \Ha & 0.077   \hspace{0.85cm} 2 \\ 
104992 & 3 & 53.1180611 & -27.7403069 & 19.65 & 7067 & 36.4$\pm$5.0 & 54 & \Ha & 0.077   \hspace{0.85cm} 2 \\ 
104992 & 4 & 53.1180649 & -27.7406502 & 19.65 & 7109 & 24.6$\pm$5.6 & 23 & \Ha & 0.077   \hspace{0.85cm} 2 \\ 
104992 & 5 & 53.1178970 & -27.7405090 & 19.65 & 7055 & 27.0$\pm$5.3 & 34 & \Ha & 0.077   \hspace{0.85cm} 2 \\ 
105723 & 1 & 53.1138153 & -27.7412968 & 20.08 & 8034 & 96.1$\pm$8.9 & 83 & \Ha & 0.223   \hspace{0.85cm} 2 \\ 
105723 & 2 & 53.1135063 & -27.7413902 & 20.08 & 8028 & 31.2$\pm$5.5 & 44 & \Ha & 0.223   \hspace{0.85cm} 2 \\ 
105723 & 3 & 53.1136093 & -27.7413673 & 20.08 & 8078 & 40.1$\pm$7.0 & 43 & \Ha & 0.223   \hspace{0.85cm} 2 \\ 
106136 & 1 & 53.1138039 & -27.7442055 & 24.40 & 6668 & 28.5$\pm$3.5 & 272 & \OIII & 0.336   \hspace{0.85cm} 1 \\ 
106136 & 1 & 53.1138039 & -27.7442055 & 24.40 & 8766 & 13.1$\pm$4.4 & 211 & \Ha & 0.336   \hspace{0.85cm} 1 \\ 
106491 & 1 & 53.1136703 & -27.7437534 & 25.04 & 6673 & 39.0$\pm$1.8 & 251 & \OIII & 0.337   \hspace{0.85cm} 1 \\ 
106491 & 1 & 53.1136703 & -27.7437534 & 25.04 & 8777 & 14.1$\pm$3.6 & 141 & \Ha & 0.337   \hspace{0.85cm} 1 \\ 
106761 & 1 & 53.1213455 & -27.7440891 & 25.81 & 8341 & 32.5$\pm$3.9 & 311 & \nodata & \nodata   \hspace{0.7cm} \nodata \\ 
108642 & 1 & 53.0944366 & -27.7341156 & 21.48 & 8673 & 91.4$\pm$9.2 & 86 & \Ha & 0.313   \hspace{0.85cm} 1 \\ 
108642 & 1 & 53.0944366 & -27.7341156 & 21.48 & 6569 & 133.5$\pm$10.9 & 91 & \OIII & 0.313   \hspace{0.85cm} 1 \\ 
108642 & 2 & 53.0944328 & -27.7341805 & 21.48 & 6524 & 141.4$\pm$11.2 & 80 & \OIII & 0.313   \hspace{0.85cm} 1 \\ 
108642 & 2 & 53.0944328 & -27.7341805 & 21.48 & 8617 & 101.1$\pm$9.9 & 81 & \Ha & 0.313   \hspace{0.85cm} 1 \\ 
109332 & 1 & 53.0898247 & -27.7366867 & 23.12 & 8958 & 18.2$\pm$4.3 & 35 & \Ha & 0.365   \hspace{0.85cm} 1 \\ 
109332 & 1 & 53.0898247 & -27.7366867 & 23.12 & 6803 & 55.3$\pm$4.0 & 56 & \OIII & 0.365   \hspace{0.85cm} 1 \\ 
109547 & 1 & 53.0892677 & -27.7360077 & 24.96 & 6809 & 38.2$\pm$1.7 & 322 & \OIII & 0.368   \hspace{0.85cm} 1 \\ 
109547 & 1 & 53.0892677 & -27.7360077 & 24.96 & 8978 & 18.8$\pm$4.3 & 219 & \Ha & 0.368   \hspace{0.85cm} 1 \\ 
109652 & 1 & 53.0903625 & -27.7367249 & 21.64 & 6800 & 18.9$\pm$1.1 & 89 & \OIII & 0.368   \hspace{0.85cm} 1 \\ 
109652 & 1 & 53.0903625 & -27.7367249 & 21.64 & 8975 & 18.5$\pm$4.9 & 186 & \Ha & 0.368   \hspace{0.85cm} 1 \\ 
109900 & 1 & 53.1128082 & -27.7346249 & 22.42 & 6499 & 11.8$\pm$3.9 & 27 & \OII & 0.744   \hspace{0.85cm} 1 \\ 
109900 & 1 & 53.1128082 & -27.7346249 & 22.42 & 8731 & 10.1$\pm$1.4 & 29 & \OIII & 0.744   \hspace{0.85cm} 1 \\ 
109953 & 1 & 53.0901070 & -27.7361164 & 21.64 & 6811 & 19.3$\pm$2.8 & 144 & \nodata & \nodata   \hspace{0.7cm} \nodata \\ 
110085 & 1 & 53.1391296 & -27.7303295 & 20.00 & 7265 & 159.7$\pm$13.6 & 80 & \Ha & 0.107   \hspace{0.85cm} 3 \\ 
110494 & 1 & 53.1079712 & -27.7337646 & 21.99 & 6358 & 121.6$\pm$10.6 & 78 & \OIII & 0.281   \hspace{0.85cm} 1 \\ 
110494 & 1 & 53.1079712 & -27.7337646 & 21.99 & 8406 & 80.3$\pm$7.5 & 76 & \Ha & 0.281   \hspace{0.85cm} 1 \\ 
111285 & 1 & 53.0815697 & -27.7334805 & 26.18 & 8897 & 50.1$\pm$7.0 & 285 & \nodata & \nodata   \hspace{0.7cm} \nodata \\ 
111549 & 1 & 53.1024857 & -27.7296772 & 22.12 & 6461 & 31.2$\pm$8.2 & 49 & \OIII & 0.314   \hspace{0.85cm} 1 \\ 
111549 & 1 & 53.1024857 & -27.7296772 & 22.12 & 8625 & 48.0$\pm$5.9 & 73 & \Ha & 0.314   \hspace{0.85cm} 1 \\ 
112157 & 1 & 53.0659561 & -27.7309017 & 24.45 & 6853 & 247.5$\pm$1.0 & 508 & \OIII & 0.376   \hspace{0.85cm} 1 \\ 
112157 & 1 & 53.0659561 & -27.7309017 & 24.45 & 9030 & 75.6$\pm$5.5 & 390 & \Ha & 0.376   \hspace{0.85cm} 1 \\ 
113173 & 1 & 53.1425781 & -27.7288322 & 24.65 & 7284 & 26.2$\pm$2.7 & 125 & \nodata & \nodata   \hspace{0.7cm} \nodata \\ 
114392 & 1 & 53.0956268 & -27.7258739 & 23.69 & 7829 & 15.9$\pm$0.9 & 52 & \OIII & 0.567   \hspace{0.85cm} 3 \\ 
116191 & 1 & 53.1104584 & -27.7176895 & 20.86 & 8008 & 63.9$\pm$5.7 & 100 & \nodata & \nodata   \hspace{0.7cm} \nodata \\ 
117070 & 1 & 53.0580940 & -27.7200108 & 23.26 & 8404 & 35.4$\pm$3.0 & 269 & \OIII & 0.683   \hspace{0.85cm} 2 \\ 
117138 & 1 & 53.0722466 & -27.7189407 & 21.24 & 8213 & 31.0$\pm$4.8 & 79 & \nodata & \nodata   \hspace{0.7cm} \nodata \\ 
117429 & 1 & 53.0714111 & -27.7176018 & 20.85 & 7589 & 24.7$\pm$3.5 & 8 & \Hb & 0.557   \hspace{0.85cm} 3 \\ 
117429 & 1 & 53.0714111 & -27.7176018 & 20.85 & 7778 & 59.9$\pm$4.5 & 19 & \OIII & 0.557   \hspace{0.85cm} 3 \\ 
117929 & 1 & 53.1232452 & -27.7181969 & 22.12 & 6668 & 13.4$\pm$2.8 & 24 & \OIII & 0.340   \hspace{0.85cm} 1 \\ 
117929 & 2 & 53.1227493 & -27.7181206 & 22.12 & 6675 & 42.3$\pm$3.5 & 66 & \OIII & 0.340   \hspace{0.85cm} 1 \\ 
117929 & 2 & 53.1227493 & -27.7181206 & 22.12 & 8794 & 15.1$\pm$3.5 & 42 & \Ha & 0.340   \hspace{0.85cm} 1 \\ 
118087 & 1 & 53.0384369 & -27.7142029 & 20.80 & 8887 & 46.7$\pm$10.9 & 30 & \Ha & 0.354  \hspace{0.85cm} 4 \\ 
118091 & 1 & 53.1171913 & -27.7188969 & 23.51 & 6685 & 30.0$\pm$3.7 & 158 & \OIII & 0.342   \hspace{0.85cm} 1 \\ 
118091 & 1 & 53.1171913 & -27.7188969 & 23.51 & 8807 & 12.7$\pm$3.8 & 110 & \Ha & 0.342   \hspace{0.85cm} 1 \\ 
118100 & 1 & 53.0703316 & -27.7179108 & 23.27 & 8224 & 41.3$\pm$1.8 & 119 & \OIII & 0.646   \hspace{0.85cm} 2 \\ 
118138 & 1 & 53.1340027 & -27.7167816 & 21.13 & 7970 & 17.7$\pm$3.2 & 19 & \nodata & \nodata   \hspace{0.7cm} \nodata \\ 
118526 & 1 & 53.0735359 & -27.7174034 & 22.67 & 8130 & 26.0$\pm$1.5 & 58 & \OIII & 0.628   \hspace{0.85cm} 3 \\ 
118673 & 1 & 53.0913582 & -27.7176514 & 24.62 & 8670 & 26.4$\pm$4.3 & 165 & \nodata & \nodata   \hspace{0.7cm} \nodata \\ 
119193\tablenotemark{\ddag} & 1 & 53.0577431 & -27.7136040 & 21.23 & 8661 & 83.0$\pm$6.9 & 81 & \OIII & 0.734   \hspace{0.85cm} 2 \\ 
119341 & 1 & 53.0700302 & -27.7166138 & 25.23 & 8222 & 4.8$\pm$1.0 & 54 & \Hb & 0.691   \hspace{0.85cm} 3 \\ 
119341 & 1 & 53.0700302 & -27.7166138 & 25.23 & 8449 & 34.5$\pm$1.3 & 315 & \OIII & 0.691   \hspace{0.85cm} 3 \\ 
119489\tablenotemark{\dag} & 1 & 53.0551300 & -27.7113838 & 19.83 & 7933 & 396.6$\pm$24.7 & 115 & \nodata & \nodata   \hspace{0.7cm} \nodata \\ 
119996 & 1 & 53.0351372 & -27.7130699 & 22.72 & 7007 & 44.3$\pm$3.5 & 122 & \nodata & \nodata   \hspace{0.7cm} \nodata \\ 
120803 & 1 & 53.0748215 & -27.7127495 & 22.73 & 8017 & 33.9$\pm$2.3 & 140 & \OIII & 0.605   \hspace{0.85cm} 3 \\ 
121127 & 1 & 53.0573578 & -27.7133713 & 21.23 & 7588 & 15.8$\pm$1.4 & 141 & \OIII & 0.519   \hspace{0.85cm} 3 \\ 
121733 & 1 & 53.0399170 & -27.7116451 & 23.49 & 6270 & 10.3$\pm$3.7 & 22 & \OII & 0.683   \hspace{0.85cm} 1 \\ 
121733 & 1 & 53.0399170 & -27.7116451 & 23.49 & 8404 & 30.6$\pm$3.9 & 88 & \OIII & 0.683   \hspace{0.85cm} 1 \\ 
121817 & 1 & 53.0965118 & -27.7111111 & 24.60 & 8345 & 32.8$\pm$3.9 & 254 & \OIII & 0.671   \hspace{0.85cm} 3 \\ 
121821 & 1 & 53.0656967 & -27.7118549 & 23.80 & 8017 & 20.6$\pm$3.7 & 112 & \nodata & \nodata   \hspace{0.7cm} \nodata \\ 
122206 & 1 & 53.0856705 & -27.7113590 & 24.76 & 7443 & 8.7$\pm$2.4 & 80 & \OII & 0.997   \hspace{0.85cm} 1 \\ 
122206 & 1 & 53.0856705 & -27.7113590 & 24.76 & 9975 & \nodata & \nodata & \OIII & 0.997   \hspace{0.85cm} 1 \\ 
122668 & 1 & 53.0660706 & -27.7097225 & 23.07 & 6242 & 8.1$\pm$1.3 & 28 & \OIII & 0.251   \hspace{0.85cm} 1 \\ 
122668 & 1 & 53.0660706 & -27.7097225 & 23.07 & 8211 & 13.5$\pm$4.4 & 61 & \Ha & 0.251   \hspace{0.85cm} 1 \\ 
123008 & 1 & 53.0693550 & -27.7090988 & 23.22 & 7975 & 8.3$\pm$1.1 & 24 & \Hb & 0.640   \hspace{0.85cm} 3 \\ 
123008 & 1 & 53.0693550 & -27.7090988 & 23.22 & 8191 & 77.0$\pm$2.4 & 217 & \OIII & 0.640   \hspace{0.85cm} 3 \\ 
123301 & 1 & 53.0775757 & -27.7081566 & 22.56 & 7800 & 11.6$\pm$1.0 & 42 & \Hb & 0.604   \hspace{0.85cm} 3 \\ 
123301 & 1 & 53.0775757 & -27.7081566 & 22.56 & 8013 & 93.4$\pm$2.3 & 285 & \OIII & 0.604   \hspace{0.85cm} 3 \\ 
123301 & 2 & 53.0772972 & -27.7082272 & 22.56 & 8001 & 134.3$\pm$2.4 & 242 & \OIII & 0.602   \hspace{0.85cm} 3 \\ 
123448\tablenotemark{\dag} & 1 & 53.0586624 & -27.7084484 & 24.71 & 8520 & 12.8$\pm$3.8 & 32 & \nodata & \nodata   \hspace{0.7cm} \nodata \\ 
123859 & 1 & 53.0642700 & -27.7057590 & 22.62 & 6888 & 3.6$\pm$0.9 & 14 & \Hb & 0.418   \hspace{0.85cm} 1 \\ 
123859 & 1 & 53.0642700 & -27.7057590 & 22.62 & 7077 & 22.0$\pm$1.2 & 82 & \OIII & 0.418   \hspace{0.85cm} 1 \\ 
123859 & 1 & 53.0642700 & -27.7057590 & 22.62 & 9338 & 17.7$\pm$6.9 & 107 & \Ha & 0.418   \hspace{0.85cm} 1 \\ 
123859 & 2 & 53.0643845 & -27.7058506 & 22.62 & 7083 & 16.1$\pm$1.1 & 73 & \OIII & 0.418   \hspace{0.85cm} 1 \\ 
124708 & 1 & 53.0647736 & -27.7055855 & 24.71 & 7075 & 22.6$\pm$4.1 & 295 & \nodata & \nodata   \hspace{0.7cm} \nodata \\ 
124761 & 1 & 53.0544395 & -27.7015305 & 21.29 & 7127 & 8.1$\pm$0.7 & 23 & \OIII & 0.427   \hspace{0.85cm} 1 \\ 
124761 & 1 & 53.0544395 & -27.7015305 & 21.29 & 9400 & \nodata & \nodata & \Ha & 0.427   \hspace{0.85cm} 1 \\ 
125541 & 1 & 53.0909691 & -27.7038784 & 23.88 & 7551 & 10.9$\pm$3.0 & 31 & \OII & 1.026   \hspace{0.85cm} 3 \\ 
125725 & 1 & 53.0384941 & -27.6966705 & 19.88 & 8171 & 70.8$\pm$10.3 & 28 & \nodata & \nodata   \hspace{0.7cm} \nodata \\ 
126769 & 1 & 53.0761490 & -27.7011623 & 23.04 & 6882 & 17.5$\pm$3.4 & 28 & \OII & 0.847   \hspace{0.85cm} 1 \\ 
126769 & 1 & 53.0761490 & -27.7011623 & 23.04 & 9249 & 24.6$\pm$8.2 & 40 & \OIII & 0.847   \hspace{0.85cm} 1 \\ 
127697 & 1 & 53.0613823 & -27.6981525 & 22.61 & 7040 & 8.5$\pm$1.3 & 15 & \OIII & 0.422   \hspace{0.85cm} 1 \\ 
127697 & 1 & 53.0613823 & -27.6981525 & 22.61 & 9333 & 18.0$\pm$5.4 & 29 & \Ha & 0.422   \hspace{0.85cm} 1 \\ 
128312\tablenotemark{\dag} & 1 & 53.0825119 & -27.6896687 & 18.99 & 6118 & 179.0$\pm$23.9 & 36 & \OIII & 0.233   \hspace{0.85cm} 1 \\ 
128312\tablenotemark{\dag} & 1 & 53.0825119 & -27.6896687 & 18.99 & 8094 & 367.4$\pm$35.1 & 59 & \Ha & 0.233   \hspace{0.85cm} 1 \\ 
128538 & 1 & 53.0531883 & -27.6954632 & 22.68 & 6900 & 4.0$\pm$0.8 & 21 & \Hb & 0.457   \hspace{0.85cm} 1 \\ 
128538 & 1 & 53.0531883 & -27.6954632 & 22.68 & 7083 & 24.5$\pm$1.8 & 140 & \OIII & 0.457   \hspace{0.85cm} 1 \\ 
128538 & 2 & 53.0531197 & -27.6958714 & 22.68 & 7092 & 55.7$\pm$3.5 & 193 & \OIII & 0.457   \hspace{0.85cm} 1 \\ 
128538 & 2 & 53.0531197 & -27.6958714 & 22.68 & 9348 & 21.3$\pm$5.8 & 101 & \Ha & 0.457   \hspace{0.85cm} 1 \\ 
129968 & 1 & 53.0494003 & -27.6943188 & 23.54 & 7795 & 7.2$\pm$2.3 & 25 & \Hb & 0.603   \hspace{0.85cm} 3 \\ 
129968 & 1 & 53.0494003 & -27.6943188 & 23.54 & 8007 & 90.9$\pm$1.9 & 257 & \OIII & 0.603   \hspace{0.85cm} 3 \\ 
130264 & 1 & 53.0469208 & -27.6908588 & 22.64 & 7681 & 17.0$\pm$3.2 & 18 & \nodata & \nodata   \hspace{0.7cm} \nodata \\ 
133441 & 1 & 53.0813599 & -27.6865711 & 24.98 & 9245 & 63.3$\pm$6.7 & 270 & \OIII & 0.851   \hspace{0.85cm} 2 \\

\end{longtable}

\scriptsize{NOTES: ---No data indicates measurement was not possible.  In the case of line IDs, no data indicates that no suitable line ID was found for the given input redshift.  ``Grism Redshift'' column gives re--calculated redshift based on the line identification.  ``Flag'' column gives source of input redshift used for line identification, if needed: 1---two lines visible in spectrum, no prior redshift needed; 2---single line in spectrum, line ID and grism redshift based on prior spectroscopic redshift; 3---single line in spectrum, line ID and grism redshift based on prior photometric redshift; 4---single line in spectrum, line ID and grism redshift based on prior spectrophotometric redshift (see Section 3.3).  Objects 68739--96627 are from the HUDF.

\dag CDF-S X-ray sources.  From Grogin \etal (2009, in preparation) matches to PEARS sources.

\ddag CDF-S X-ray sources with $L_{X}$\cge$10^{42}$erg $s^{-1}$ and thus likely AGN.  From Grogin \etal (2009, in preparation).
}

\vspace{1cm}

\begin{longtable}{cccccccccc}
\tabletypesize{\scriptsize}
\tablecaption{Summary of ELG Detections in South Fields \label{table2}}
\tablewidth{0pt}
\tablehead{
\colhead{Field}&\colhead{\# of}&\colhead{\# of}&\colhead{\# of}&\colhead{\# of Galaxies with}&\colhead{\# of Knots with}\\
\colhead{}&\colhead{Lines}&\colhead{Knots}&\colhead{Galaxies}&\colhead{multiple knots}&\colhead{multiple lines}\\
}

\startdata
HUDF    & 98 & 78 & 64 & 12 & 15 \\
South 1 & 51 & 37 & 35 & 2 & 8 \\
South 2 & 55 & 34 & 31 & 3 & 9 \\
South 3 & 50 & 47 & 35 & 6 & 15 \\
South 4 & 66 & 30 & 27 & 3 & 10 \\
TOTAL & 320 & 226 & 192 & 26 & 61 \\


\end{longtable}

\scriptsize{NOTE: ---Here knots with multiple lines means two lines sufficient to deduce a wavelength ratio and therefore secure grism redshift; i.e. \emph{not} \OIII and \Hb since a set wavelength ratio was used in the fitting algorithm.}

\vspace{1cm}

\begin{longtable}{ccccccccccc}
\tabletypesize{\scriptsize}
\tablecaption{Summary of Lines Detected in South Fields \label{table2}}
\tablewidth{0pt}
\tablehead{
\colhead{Field}&\colhead{\#}&\colhead{\#}&\colhead{\#}&\colhead{\#}&\colhead{\#}&\colhead{\#}&\colhead{\#}&\colhead{\#}&\colhead{\#}&\colhead{\#}\\
\colhead{}&\colhead{\OIII}&\colhead{\Ha}&\colhead{\OII}&\colhead{\Hb}&\colhead{\CIV}&\colhead{\CIII}&\colhead{MgII}&\colhead{NeIII}&\colhead{H$\gamma$}&\colhead{No ID}\\
}
\startdata
HUDF & 43 & 31 & 13 & 8 & 1 & 1 & 1 & 0 & 0 & 4 \\
South 1 & 20 & 11 & 6 & 7 & 1 & 1 & 0 & 0 & 0 & 7 \\
South 2 & 26 & 10 & 4 & 8 & 1 & 1 & 1 & 0 & 0 & 3 \\
South 3 & 22 & 24 & 4 & 2 & 0 & 0 & 0 & 1 & 1 & 9 \\
South 4 & 26 & 8 & 4 & 4 & 0 & 0 & 0 & 0 & 0 & 8 \\
TOTAL & 136 & 83 & 30 & 30 & 3 & 3 & 2 & 1 & 1 & 31 \\
\end{longtable}

\end{document}